\documentclass[10pt]{article}
\usepackage[utf8]{inputenc}
\usepackage[margin=1in]{geometry}
\usepackage{natbib}
\usepackage{amsmath}
\usepackage{amssymb}
\usepackage{amsthm}
\usepackage{subfig}
\usepackage{graphicx}
\usepackage{fancyhdr}
\usepackage{hyperref}
\usepackage{tikz}
\usepackage{enumerate}
\usepackage{authblk}
\usepackage{float}
\usepackage{subfig}
\usepackage{algorithmic}

\usepackage[normalem]{ulem}

\newcommand{\cov}{\text{cov}}
\newcommand{\bs}{\boldsymbol}

\newcommand{\R}{\mathbb{R}}
\newcommand{\N}{\mathbb{N}}
\usepackage[noend,ruled]{algorithm2e}

\newcommand\norm[1]{\left\lVert#1\right\rVert}
\newcommand\normal{\mathrm{N}}

\date{}
\begin{document}
\title{Plateau Proposal Distributions for Adaptive Component-wise Multiple-Try Metropolis}
\author[1]{F. Din-Houn Lau}
\author[2,3]{Sebastian Krumscheid}
\affil[1]{
Department of Mathematics, Imperial College London, London, UK
}

\affil[2]{
  Department of Mathematics and Computer Science, Freie Universit{\"a}t Berlin, Germany}
\affil[3]{
   Department of Mathematics, RWTH Aachen University, 52062 Aachen, Germany
}

\maketitle

\begin{abstract}
  Markov chain Monte Carlo (MCMC) methods are sampling methods that
  have become a commonly used tool in statistics, for example to
  perform Monte Carlo integration. As a consequence of the increase in
  computational power, many variations of MCMC methods exist for
  generating samples from arbitrary, possibly complex, target
  distributions. The performance of an MCMC method is predominately
  governed by the choice of the so-called proposal distribution
  used. In this paper, we introduce a new type of proposal
  distribution for the use in MCMC methods that operates
  component-wise and with multiple trials per iteration. Specifically,
  the novel class of proposal distributions, called \textit{Plateau}
  distributions, do not overlap, thus ensuring that the multiple
  trials are drawn from different regions of the state
  space. Furthermore, the Plateau proposal distributions allow for a
  bespoke adaptation procedure that lends itself to a Markov chain
  with efficient problem dependent state space exploration and
  improved burn-in properties. Simulation studies show that our novel
  MCMC algorithm outperforms competitors when sampling from
  distributions with a complex shape, highly correlated components or
  multiple modes.
\end{abstract}

\section{Introduction}

Markov chain Monte Carlo (MCMC) methods are essentially used to
perform Monte Carlo integration, which has become a standard
statistical tool. Specifically, MCMC methods produce samples from a
target distribution $\pi$ by using an ergodic Markov chain with
stationary distribution $\pi$. Typically, MCMC methods are used when
it is difficult to sample from the target distribution directly, e.g.\
when the normalisation constant is unknown. There are many ways to
construct this Markov chain which have lead to many variations of MCMC
methods; see, e.g., \cite{brooks2011handbookmcmc}.

The classic MCMC method is the Metropolis-Hastings algorithm
\citep{annealing2}. At each iteration, the Metropolis-Hastings
algorithm is designed to update the entire current state (i.e.\ all
components of the random vector generated at the previous iteration)
at once. However, updating individual components, or subsets of
components, is possible. Indeed, this type of component-wise updating
was initially proposed in \cite{annealing2}, but did not receive much
attention at first. In this paper, we focus on updating individual
components. Using individual component updates is a way of sampling
from the (lower dimensional) conditional distributions of the target,
provided the conditional distributions are known. However, this is
typically not the case in practise. An approach to remedy this is to
model the conditional distributions. However, using parametric models
leads to an inflexible approximation whereas non-parametric models do
not scale well with the number of dimensions. For these reasons, we
focus our attention on independent component-wise updates.

Another variant of MCMC sampling is the multiple-try method
\citep{multipletryMCMC} where several proposals or \textit{trials} are
suggested at each iteration, as opposed to a single proposal. The
motivation behind the multiple-try method is that more of the space is
explored at the expense of an increased computational cost (i.e.\
proposal generation and evaluation of acceptance criterion). In fact,
the multiple-try approach is known to offer great flexibility in terms
of designing MCMC methods that satisfy detailed balance. As a
consequence, several variations exist in the literature; see also the
discussion in~\cite{MR3141364}.

Recently, in \cite{doi:10.1080/10618600.2018.1513365} the authors
introduce a component-wise, multiple-try MCMC method with Gaussian
proposals for each trial, where each univariate proposal has a
different variance. In this work, we introduce a new class of proposal
distributions for use in a component-wise, multiple-try MCMC
method. These proposals, called \textit{Plateau} distributions, do not
overlap to exploit the multiple-try nature of the method. Indeed, by
using proposals that do not overlap for each trial, the Markov chain
is forced to explore different parts of the state-space. Conversely,
using proposals that overlap, e.g.\ Gaussians with different
variances, can lead to an inefficient algorithm, as the trials tend to
be from a similar region of the state-space.

Using Plateau distributions for the proposal distributions leads to
better exploration of the space over standard Gaussian proposals. The
idea is intuitive, easy to implement and leads to good results, in the
sense of exploring the state space. Moreover, using the Plateau
proposals leads to a reversible Markov chain with the target
distribution as its invariant distribution e.g.\ see
\cite{doi:10.1080/10618600.2018.1513365}.

As is common for most proposals used in MCMC methods, the Plateau
distributions depend on parameters that need to be selected with
respect to the target distribution to obtain an effective MCMC
algorithm.  Adaptation of MCMC methods typically entails
\textit{tuning} the parameter(s) of a class of proposal distributions,
e.g.\ the variance parameter in a Normal distribution, in order to
improve the convergence properties of the Markov chain. For instance,
in \cite{haario2001adaptive}, a Gaussian proposal distribution is used
whose variance (or covariance) is adapted using the previously
generated states of the Markov chain.

\begin{figure}[bp]
  \centering
\subfloat[][]{\includegraphics[width=0.49\linewidth]{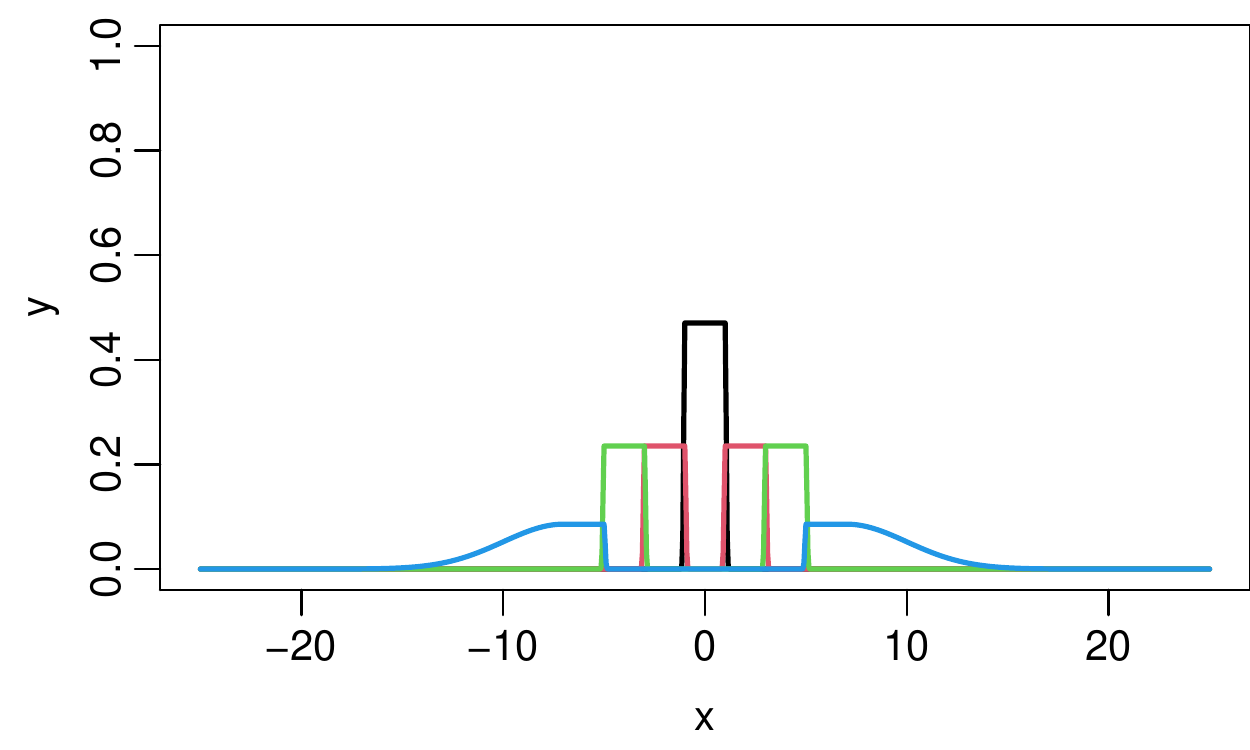}\label{fig:ill1}}

\subfloat[][]{\includegraphics[width=0.49\linewidth]{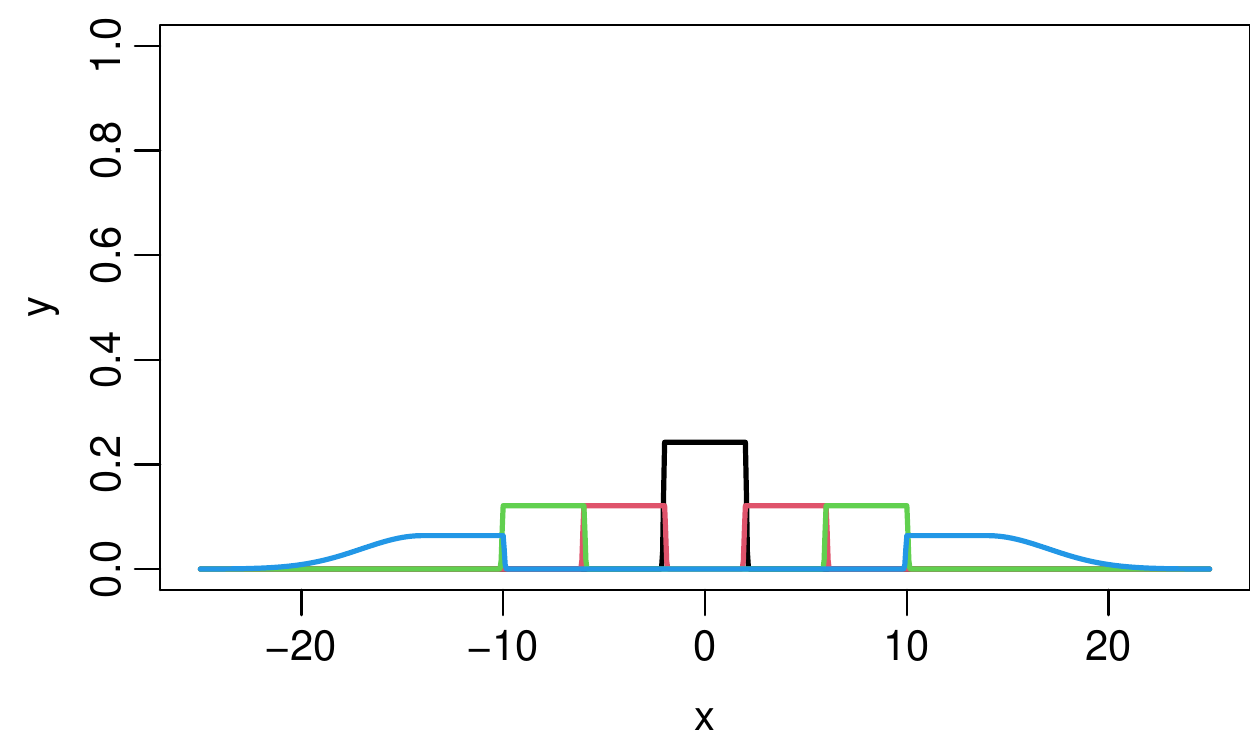}\label{fig:ill2}}
\subfloat[][]{\includegraphics[width=0.49\linewidth]{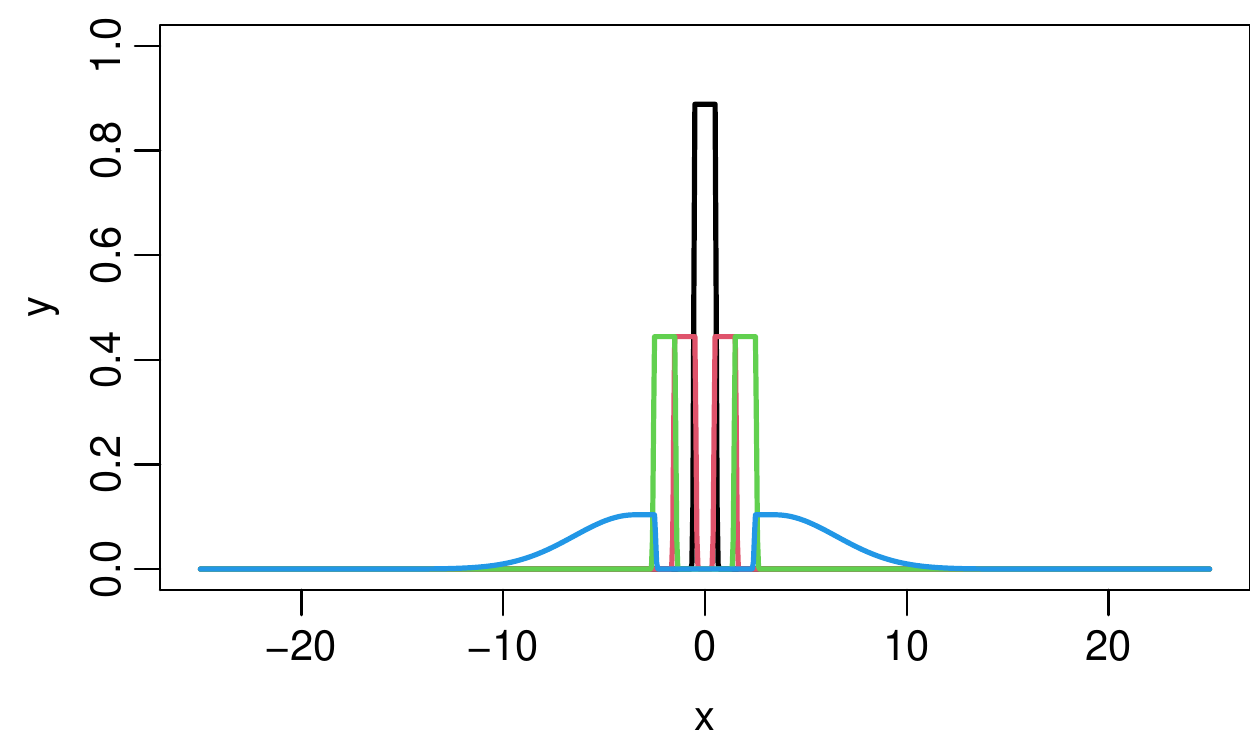}\label{fig:ill3}}

\caption{Illustration of $4$ Plateau proposal distributions (probability density functions)
  represented by the different colours and
  the adaptation procedure. The initial Plateau distributions are
  presented in (a); If the outermost (blue) proposals are
  selected frequently, the proposals are adapted to (b); If innermost
  (black) proposals are
  selected frequently, the proposals are adapted to (c). 
  }
  \label{fig:illus}
\end{figure}

In this work we propose an adaptation procedure that is designed for
use with the non-overlapping multiple-try Plateau proposals. The
Plateau proposals together with the associated adaptation procedure
are exemplified in Fig.~\ref{fig:illus}. Suppose the MCMC algorithm is
initiated with multiple-try proposals whose distributions are
presented by the different coloured lines in
Fig.~\ref{fig:ill1}. These proposals operate independently on each
(univariate) component in the target space. As the Markov chain
evolves, the number of times each Plateau's proposed candidate state
is selected is recorded. If the innermost or outermost Plateaus are
overly selected, then the set of Plateau proposals are re-organised as
depicted in Figs.~\ref{fig:ill2} and \ref{fig:ill2}. More precisely,
if the innermost (black) proposal is overly selected, then one halves
the width of each plateau and shifts the set of proposals to remove
the gaps. A similar procedure is conducted if the outermost (blue)
proposals is overly selected by doubling the width of the
plateaus. This procedure appropriately scales the set of Plateau
proposals to the target distribution. Notice that this intuitive
adaptation procedure makes explicit use of the non-overlapping feature
of the Plateau proposals. Specifically, if a sample is selected from a
particular Plateau distribution, then that sample could not have been
obtained by sampling from any other Plateau distribution since their
supports do not overlap. Therefore, counting the number of times each
Plateau is selected gives a clear and direct indication as to how to
update the Plateaus to match the shape of the target. This indication
is not so clear when using proposals that overlap. The advantages of
using non-overlapping proposals, as opposed to overlapping proposals,
is explored in Section \ref{sec:invest-adapt-mcmc}. The mathematical
definition of the Plateau proposals and the pseudocode for the
adaptation procedure are presented later in Sections
\ref{sec:non-overl-prop} and \ref{sec:adaptation-proposals}
respectively.

The works in \cite{MR2758304} and \cite{MR2380640} share a similar
objective with our proposed work, namely: draw samples that are far
apart from each other to facilitate an effective exploration of the
state-space. For example, in \cite{MR2758304} the authors introduce an
independent (i.e., component-wise), yet single-try,
Metropolis--Hastings method using a normal mixture distribution as
proposal. To achieve a more structured state space exploration, the
mixture distribution is adapted, based an appropriate $k$-means
clustering of chain's visited states.  In \cite{MR2380640} an improved
state space exploration is achieved by combining variance reduction
techniques with multiple trials. Specifically, it involves drawing
multiple trails from a single, well-chosen proposal distribution that
is based on, e.g., Latin Hypercube sampling. The performance of these
approaches depends on a well-chosen transformation. Conversely, the
approach introduced in this work does not require such a
transformation, as the Plateau proposals reside in the state space.
Further, our adaptation procedure automates the tuning of the Plateau
proposals parameters and does not require additional clustering (i.e.,
optimisation) algorithms to be performed.

To summarise, the goal of this work is to propose a new MCMC algorithm
along with an adaptation procedure, which is simple to implement,
intuitive and can by used without using explicit derivative
information (e.g.\ Hamiltonian Monte Carlo \cite{neal2011mcmc}) or
conditional distributions (e.g.\ Gibbs sampling \cite{gibbs92}) of the
target. Therefore a general-purpose MCMC algorithm is required. To
this end, we introduce Plateau proposals as a general class of
proposal distributions for use in a component-wise multiple-try MCMC
methods. Specifically, the combination of non-overlapping
characteristic of Plateau proposals with the multiple-try approach
lends itself to a bespoke adaptation procedure that leads to good
results for a variety of target distributions. This is achieved by
improved state space exploration and favourable burn-in properties.

  The remainder of this is organised as follows. In Section
  \ref{sec:gener-proc-comp} a generic component-wise multiple-try
  algorithm is presented. The novel class of Plateau proposals is
  introduced and discussed in Section \ref{sec:non-overl-prop}. In
  Section \ref{sec:adaptation-proposals} we discuss how to adaptively
  select the parameter of the Plateau proposals and offer a detailed
  algorithmic description of the complete method.  The performance of
  our new method is then compared with other MCMC methods in Section
  \ref{sec:results}. Finally, a commentary on improvements and a
  summary are provided in Section \ref{sec:conclusion}.

\section{Component-wise Update with
  Multiple Trials}\label{sec:adapt-comp-wise}

Let $\pi$ be a probability density function,
$\pi\colon \mathcal{X} \to \mathbb{R}^+$, where
$\mathcal{X}\subseteq \mathbb{R}^d$. Our main interest is to sample
from $\pi$; this is the target distribution. We assume that sampling
directly from $\pi$ is difficult or impossible, for example because
$\pi$ may only be known up to a multiplicative constant. In order to
sample from $\pi$ we use MCMC methods.  Specifically, given the
current state $\bs{X}_n = \bs x \in\mathcal{X}$ of the Markov chain at
iteration $n\in\N$, the Metropolis-Hastings algorithm, one of the
simplest MCMC methods, proposes a candidate $\bs Y$ for the chain's
next state $\bs{X}_{n+1}$ by drawing a random variable from
distribution with probability density function (PDF)
$T(\bs x,\cdot)\colon\mathcal{X}\to\R^+$ from which it is easier to
sample from than from the target distribution. That is
$\bs Y \sim T(\bs x,\cdot)$, where $T(\bs x,\cdot)$ is the conditional
density given the current value $\bs x$ of, what is commonly known as,
the proposal distribution.  For example, the random-walk proposal
\cite{annealing2,HASTINGS01041970} uses a multivariate normal
distribution, which we will write as
$T(\bs x,\cdot) = \normal(\bs x,\sigma^2 I)$ with $\sigma>0$ given and
$I$ being the $d\times d$ identity matrix. Another example is
$\normal\bigl(\bs x + \tau \nabla \ln\bigl(\pi(\bs x)\bigr),2\tau I)$
with $\tau>0$ fixed, which is the proposal used in the
Metropolis-adjusted Langevin algorithm \cite{Roberts1996}. The
realisation of the Markov chain's next state is then selected by means
of an accept-reject procedure in a way such that the resulting Markov
chain's stationary distribution is $\pi$.

One of the main difficulties when using an MCMC method is the choice
of the proposal distribution. In particular, the choice of the
proposal may significantly affect the properties of the MCMC method,
including the speed of convergence to equilibrium and mixing
properties \cite{grandstrand:2004}. Typically, the proposal
distribution is selected from some family of well-known distributions,
e.g., from the family of normal distributions. It is noteworthy that
an optimal MCMC performance in high dimensions requires to select the
``scale'' of the proposal appropriately \citep[e.g.\
see][]{roberts2001}.

Instead of proposing a single multivariate candidate from
$\bs Y \sim T(\bs x, \cdot)$ by updating all components of the current
state $\bs{X}_n = \bs x$ simultaneously via the (global) proposal
distribution $T(\bs x,\cdot)$,
it is also possible to split the state space $\mathcal{X}$ into its
individual components (or small groups of components) and propose
candidates for each component (or group of components)
independently.
This local (or projected) approach is intuitive, computationally
efficient, and reduces the problem of selecting a multidimensional
proposal into lower dimensional proposals that are easier to
handle. MCMC methods using these types of updates
are called component-wise MCMC methods which may use a potentially
different proposal distribution per component of the state
\citep[Ch.~1]{gilks1995markov}. In this work we focus on
one-dimensional, component-wise proposals. We emphasise, however, that
the ideas that follow are not restricted to one-dimensional components
and are general in fact.

Considering independent one-dimensional, component-wise proposals is
equivalent to the proposal distribution $T(\bs x,\cdot)$ given
$\bs x = (x_1, \dots, x_d)\in\R^d$ being separable, in the sense that
\begin{equation}
  T(\bs x,\bs y) \equiv T_1(x_1,y_1)\times\dots\times T_d(x_d,y_d) =
  \prod_{k=1}^d T_k(x_k,y_k)\;,\label{eq:1dcompwise:proposal}
\end{equation}
for any $\bs y = (y_1, \dots, y_d)\in\R^d$. Here,
$T_k(x,\cdot)\colon \R\to\R^+$, $k=1,\dots,d$ with $x\in\mathbb{R}$,
denotes the one-dimensional proposal density given $x$ used
to draw the candidate for component $k$. That is, the proposed
(global) candidate $\bs Y = (Y_1,\dots, Y_d)\in\R^d$ is obtained by
sampling each $Y_k\sim T_k(x_k,\cdot)$ mutually independent from any
other $Y_\ell$.
Notice that this one-dimensional, component-wise proposal step is
identical to the standard MCMC method with $d=1$, and hence
can be considered a natural extension to the multivariate case.

While the component-wise candidate proposal
\eqref{eq:1dcompwise:proposal} is computationally efficient, by
construction it does not account for correlations between the
components. Consequently, the proposed candidates $\bs Y$ may not be
good representatives of the target distribution $\pi$ (i.e.\
  most candidates $\bs Y$ will be rejected, resulting in a very low
  acceptance rate), if $\pi$ has highly correlated
components. Therefore these ``uninformed'' candidates may lead to a
poor state space exploration and thus to a poor performance of the
MCMC method. To remedy these defects, we will combine component-wise
proposals with multiple trials for each component. Specifically,
the multiple-try technique works by proposing many candidates from a
proposal distribution, rather than just a single one, amongst which
the ``best'' one is selected. Each trial may be proposed from a
different proposal distribution. Thus, in combination with
component-wise proposals, we (independently) generate $M$
possible candidates (i.e., $M$ trials) independently for each
component $k=1,\dots, d$. Let $T_{j,k}(x,\cdot)\colon\R\to\R^+$,
$j=1,\dots,M$, denote the proposal PDF of the $j$th trial for the
$k$th component given $x_k=x$.

\subsection{Generic Procedure of Component-wise Multiple-Try
  Metropolis}\label{sec:gener-proc-comp}
A generic component-wise multiple-try algorithm is now described --
the full pseudo-code for generating $N$ samples from the
  target distribution $\pi$ is presented in Algorithm
\ref{alg:very_generic}. Each MCMC iteration of the algorithm
involves drawing multiple trials and then performing an
acceptance-rejection step for each component sequentially.

We now describe the intuition behind the steps of the
procedure. Suppose that the state of the chain at the beginning of the
$n$th iteration is $\bs x = (x_1,\dots,x_d)$. For the first component
(i.e., $k=1$), $M$ independent trials, $z_{1},\dots,z_{M} \in\R$, are
drawn from $T_{j,1}(x_1,\cdot)$.  These trials are then weighted
according to
  \begin{equation*}
    w_{j,1}(z_{j},\bs x ) = \pi\bigl((z_{j}; \bs x_{[-1]})\bigr) T_{j,1}(x_1,z_{j})
  \lambda_{j,1}(x_1,z_{j})\;,
\end{equation*}
where $(z; \bs x_{[-i]}) \in \R^d$ denotes the vector that is
  identical to $\bs x$ except for its $i$th component which
  is replaced by $z\in \R$, that is
$(z; \bs x_{[-i]}) = (x_1,\dots,x_{i-1}, z, x_{i+1},\dots,
x_d)$.

The functions $\lambda_{j,k}(x,y)$ with $x,y\in\mathbb{R}$ for any
$k=1,\dots d$, are non-negative, symmetric functions in $x$ and $y$
which are selected by the user. Further it is required that
$\lambda_{j,k}(x,y)>0$ whenever $T_{j,k}(x,y)>0$. Each trial $z_{j}$,
$j=1,\dots,M$, has an associated weight $w_{j,1}(z_{j},\bs x )$. A
candidate for the first component of the chain's next state is then
randomly selected amongst all trials $z_{j}$ $(j=1,\dots,M)$ according
to these weights. The selected candidate is then accepted or
rejected. 
The remaining components $k=2,3,\dots, d$ of the chain's state are
updated in order in a similar fashion; Algorithm
\ref{alg:very_generic} illustrates a detailed pseudo-code of the
corresponding MCMC method.

\renewcommand{\algorithmicrequire}{\textbf{Input:}}
\renewcommand{\algorithmicensure}{\textbf{Output:}}

\begin{algorithm}[tp]
  \caption{Generic Component-wise Multiple-Try Metropolis}\label{alg:very_generic}
  \begin{algorithmic}[1]
    \REQUIRE number of trials $M$; number of MCMC realisations
    $N$; 
    starting position $\bs x_0\in\mathbb{R}^d$; target distribution
    $\pi$ (possibly un-normalised); proposal distributions $T_{j,k}$
    \STATE Let $\bs X_0=\bs x_0 = \bs x$.

    \FOR{$n=1,\dots,N$}
    \FOR{$k=1,\dots,d$}
    \STATE   Propose $M$ trials: $z_{j} \sim T_{j,k}(x_k,\cdot)$ for
    $j=1,\dots,M$.
    \STATE  Compute the trial weights
    \begin{equation*}
      w_{j,k}(z_{j},\bs x ) = \pi\bigl((z_{j}; \bs x_{[-k]})\bigr) T_{j,k}(x_k,z_{j})
      \lambda_{j,k}(x_k,z_{j})\;,\quad j=1,\dots,M.
    \end{equation*}
    \STATE Draw $y \in \{z_{1},\dots, z_{M} \}$
      randomly 
    with probability proportional to $w_{1,k},\dots,w_{M,k}$.  \STATE
    Draw $x_{j}^\ast\sim T_{j,k}( y , \cdot)$ for $j=1,\dots,M-1$ and
    let $x_M^\ast = x_k$.  \STATE Let $\bs y = (y;\bs x_{[-k]})$ and
    compute
    \begin{equation*}
      \alpha = \min\left\{1 , \frac{w_{1,k}(z_{1},\bs x) + \dots +
          w_{M,k}(z_{M},\bs x)  }{w_{1,k}(x^\ast_1,\bs y) + \dots +
          w_{M,k}(x^\ast_M,\bs y)}  \right\}.
    \end{equation*}
    Draw $r\sim \text{Uniform}(0,1)$.
    \IF{$r < \alpha$}
    \STATE   Accept $\bs X_n = \bs y = (y;\bs x_{[-k]})$ and set $\bs x =
    \bs y$.
    \ELSE{ }
    \STATE $\bs X_n = \bs x$.
    \ENDIF
    \ENDFOR
    \ENDFOR

    \RETURN $\bs X_1,\dots, \bs X_N$
  \end{algorithmic}
\end{algorithm}

Specifically, to propose a candidate for the $k$th component of the
Markov chain's next state in step~5 of
Algorithm~\ref{alg:very_generic} each trial for the current $k$th
component 
$x_k$ has an associated weight
\begin{equation*}
 w_{j,k}(z_{j},\bs x) = \pi\bigl((z_{j}; \bs x_{[-k]})\bigr) T_{j,k}(x_k,z_{j})
  \lambda_{j,k}(x_k,z_{j}),\quad j=1,\dots,M.
\end{equation*}
As described above, a single trial is then randomly selected with
probabilities proportional to $w_{j,k}$. There are a number of choices
for $\lambda_{j,k}$ in the multiple-try literature
\cite{multipletryMCMC}, such as
\begin{equation*}
  \left(\frac{T_{j,k}(x,y) + T_{j,k}(y,x)}{2}\right)^{-1},\quad
  \left\{T_{j,k}(x,y)T_{j,k}(y,x)\right\}^{-\beta} \quad\text{and}\quad 1.
\end{equation*}
In work \cite{doi:10.1080/10618600.2018.1513365} the authors
  suggest to use
\begin{equation}\label{eq:lambda}
  \lambda_{j,k}(x,y) = T_{j,k}(x,y)\norm{y-x}^{\alpha}\;,
\end{equation}
where $\alpha = 2.9$ was used following a simulation study focusing on
large moves in the state spaces. Here we will also use
  $\lambda_{j,k}$ as in \eqref{eq:lambda}. However, in simulations,
not reported here, we found that $\alpha = 2.5$ performed best in
terms of the mean squared error for a variety of target
distributions. Consequently we use $\lambda_{j,k}$ with $\alpha = 2.5$
for the remainder of the paper. We note that it is beyond the scope of
this paper to propose a particular form for the class of functions
$\lambda_{j,k}$ due to the problem dependent nature of this
choice. Instead we advocate that users perform a trial run with a
variety of $\lambda_{j,k}$ to determine which is best suit for their
application and performance metric.

\section{Non-Overlapping Proposal Distributions}\label{sec:non-overl-prop}
In principle, any family of proposals $T_{j,k}$ can be used in
Algorithm \ref{alg:very_generic}. However, a careful choice of the
proposals can lead to a more efficient
algorithm. 

\begin{figure}[tbp]
  \centering
  \includegraphics[width=0.5\linewidth]{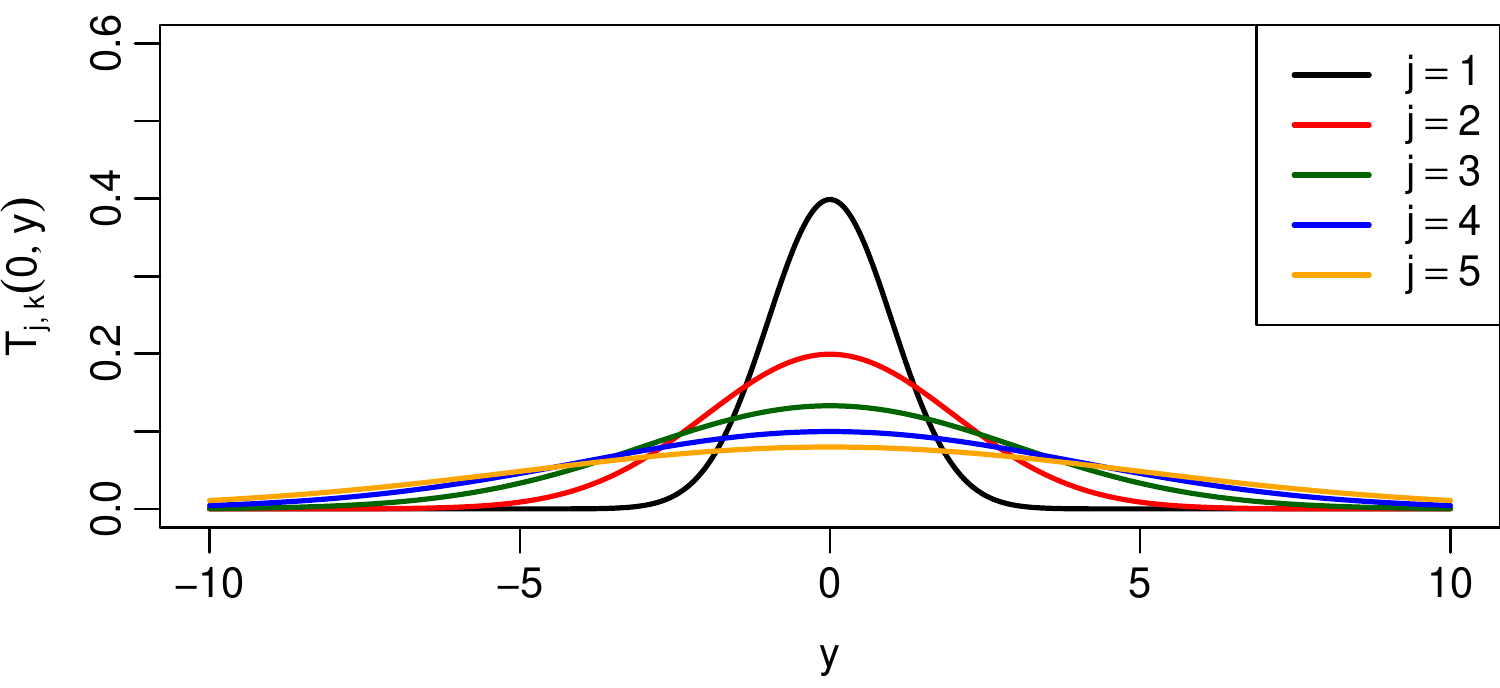}
  \caption{Probability density function of 5 Normal distributions with
    zero mean and standard deviations $j\sigma$ for
      $\sigma=1$, $j=1,\dots,5$. }
  \label{fig:p1_pdf}
\end{figure}

Recall that $T_{j,k}$ is the $j$th trial proposal distribution for the
$k$th component. As mentioned earlier, the motivation of using
multiple trials is to explore a larger region of the state space than
is achieved by using a single proposal. Therefore, it would not be
beneficial to use proposals that are similar. To illustrate this,
suppose that for a fixed component $k$ we use the proposal
  distributions $T_{j,k}(x,\cdot)=\normal(x ,(j\sigma)^2)$ for
$j=1,\dots,M=5$ with known $\sigma>0$ and take $x=0$ without loss of
generality. The probability density functions of these proposals of
with $\sigma = 1$ are presented in Figure \ref{fig:p1_pdf}.

As illustrated in Figure \ref{fig:p1_pdf}, these proposals are very
similar. Indeed, $99$\% of $T_{1,k}$'s density mass lies within the
interval
\begin{equation*}
  \mathfrak{J} = \bigl( -\sigma \Phi^{-1}(0.995), \sigma \Phi^{-1}(0.995)\bigr)
 \end{equation*}
 centred around $x=0$, where $\Phi^{-1}$ is the inverse of the
 cumulative distribution function of a standard normal so that
 $\Phi^{-1}(0.995)\approx 2.6$.  Suppose that we consider for the second
 proposal distribution $Y_{2,k}\sim T_{2,k}(0,\cdot)$, then
 $\mathbb{P}(Y_{2,k}\in \mathfrak{J}) \approx 0.8$ for any $\sigma>0$. That is, draws
 from $T_{1,k}$ and $T_{2,k}$ will be located in the same region with
 high probability. Similar arguments hold for the wider Gaussian
 proposals.  Thus draws from these Gaussian proposals will tend to be
 similar, thus leading to an inefficient use of the multiple-try
 technique. To avoid sampling similar proposals for each trial, we
 seek densities which do not overlap (or overlap to a small
 degree).
 
 \begin{figure}[tbp]
  \centering
\subfloat[][]{  \includegraphics[width=0.49\linewidth]{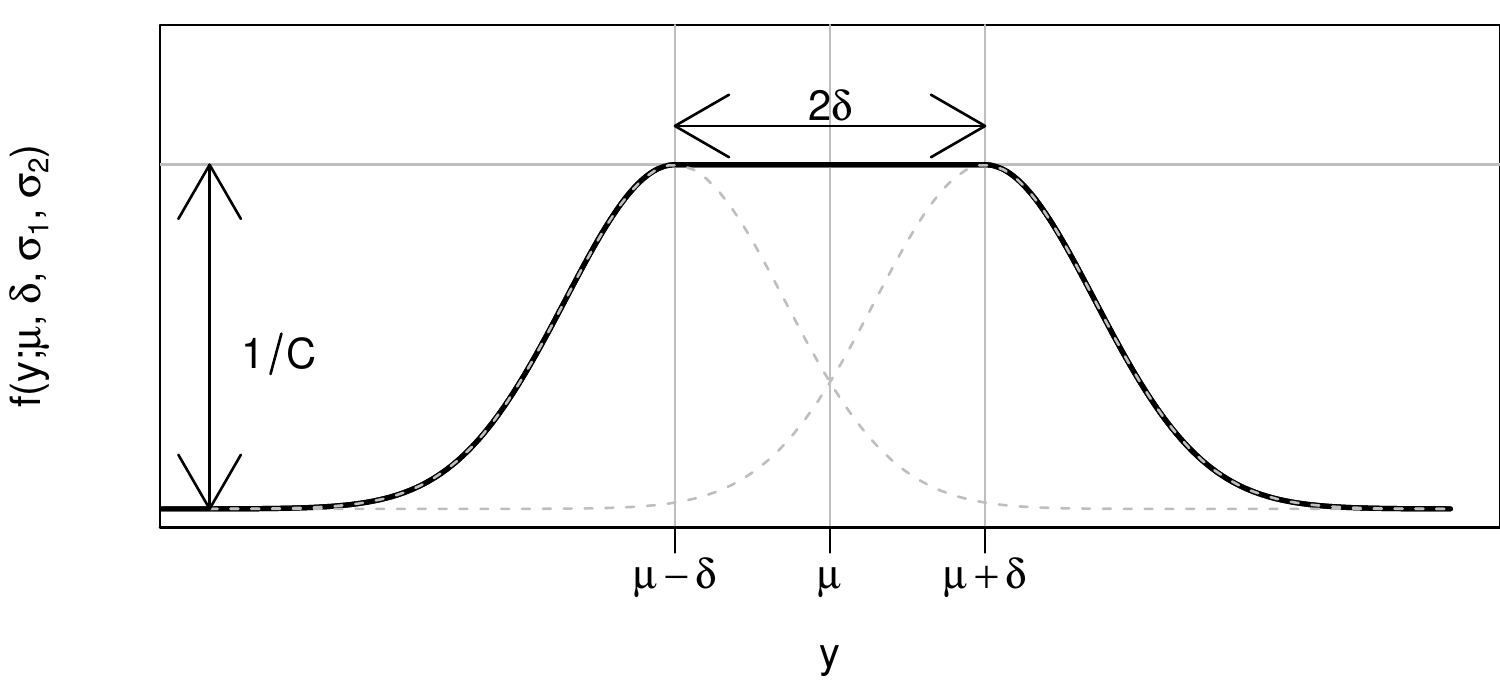}\label{fpdf}}
\subfloat[][]{    \includegraphics[width=0.49\linewidth]{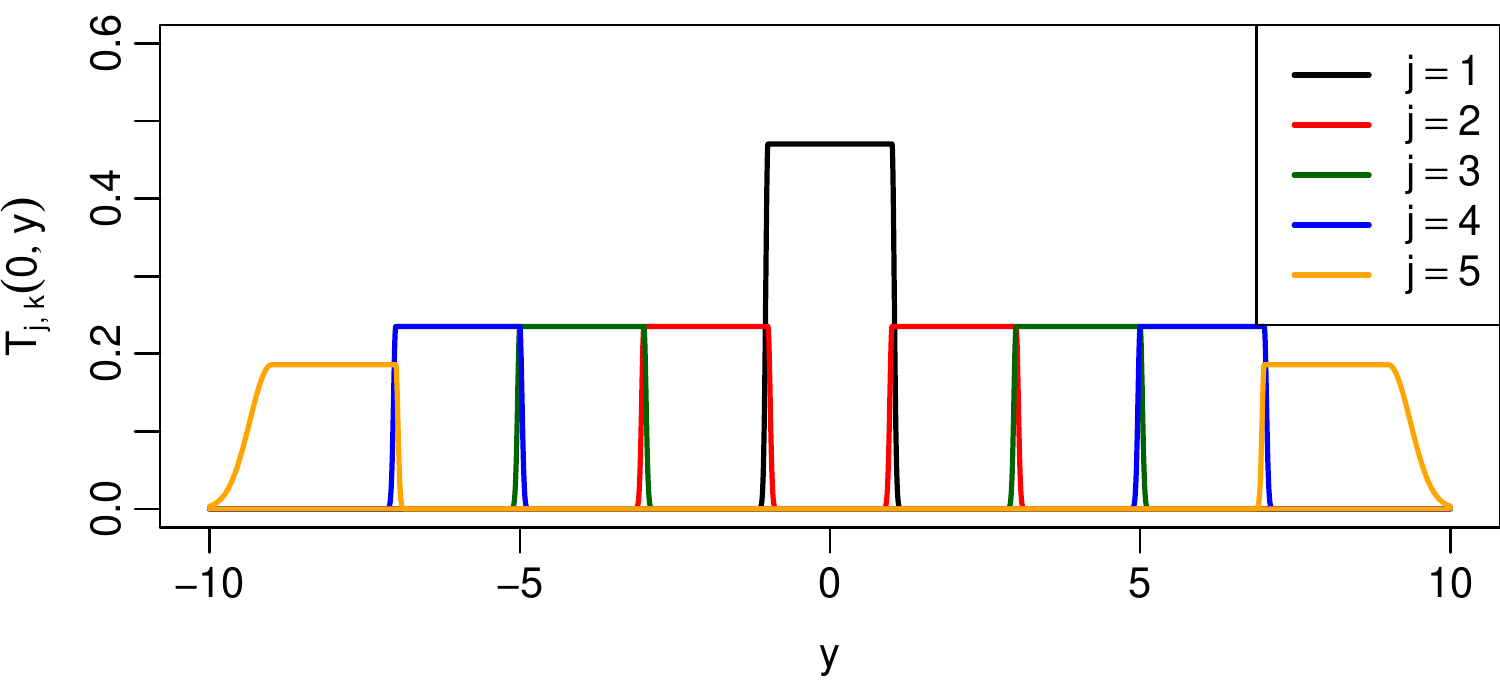}\label{fig:p2_pdf}}

  \caption{PDF of Plateau proposal distributions with $M=5$. The five
    different proposals each have a different colour. 
  }
  \label{fig:p2andf}
\end{figure}
Specifically, we advocate using proposals of the type illustrated in
Figure \ref{fig:p2_pdf}.  That is, each trial for each component-wise
proposal distribution combines uniform distributions with
exponentially decaying tails. Notice, that this means that the amount
of overlap between different proposals 
is controlled through how fast the tails
decay. 
Specifically, we first introduce the
  PDF 
\begin{equation*}
  f(y; \mu,\delta,\sigma_1,\sigma_2)= \frac{1}{C}
  \begin{cases}
    \exp\left\{-\frac{1}{2\sigma_1^2}\left[y-(\mu-\delta) \right]^2 \right\}& \text{for }y< \mu-\delta\\
    1 &  \text{for }\mu-\delta\leq y\leq \delta+\mu\\
    \exp\left\{-\frac{1}{2\sigma_2^2}\left[y-(\mu+\delta) \right]^2 \right\}&  \text{for } y> \mu+\delta\\
  \end{cases}
\end{equation*}
where
\begin{equation*}
  C = \frac{\sqrt{2\pi \sigma_1^2}}{2} + \frac{\sqrt{2\pi\sigma_2^2}}{2} + 2\delta
\end{equation*}
denotes the normalisation constant. This PDF is illustrated in Figure
\ref{fpdf}. For each component $k$ with given value $x_k = x$, we then
set the PDF of the each trial proposal as
\begin{equation*}
  T_{j,k}(x,y)=
  \begin{cases}
    f(y; x, \delta_1, \sigma, \sigma) & j=1\\
    \frac{1}{2}f(y; x - 2(j-1)\delta_1 -\delta, \delta, \sigma, \sigma) +
    \frac{1}{2}f(y;x + 2(j-1)\delta_1 + \delta, \delta, \sigma, \sigma) & j=2,\dots,M-1,\\
    \frac{1}{2}f(y; x - 2(M-1)\delta_1 - \delta, \delta, \sigma_0, \sigma) +
    \frac{1}{2}f(y; x + 2(M-1)\delta_1 + \delta, \delta, \sigma, \sigma_1) & j=M
  \end{cases}
\end{equation*}
for some values of $\delta_1, \delta, \sigma, \sigma_0, \sigma_1>0$. 
The $M=5$ trial proposals shown in Figure \ref{fig:p2_pdf}
correspond to $\delta_1 = \delta = 1$, $\sigma = 0.05$ and $\sigma_0 =
\sigma_1 = 0.5$.
We shall refer to the proposals of this type as \emph{Plateau} proposals
given the shape of their PDFs. The $\delta_1$ parameter controls the
width of the central Plateau centred at the current state $x$. The
$\delta$ parameter is the width of the other Plateaus. The $\sigma$
value controls the decay of the tails either side of the inner
Plateaus. The outer tails for the $M$th proposal are described by
$\sigma_0$ and $\sigma_1$.

To compare with the earlier calculations for coverage probabilities
for the Gaussian proposals; $99$\% of the density of $T_{1,k}$ with
$x=0$, $\delta = 1$, and $\sigma = 0.5$ lies in the interval
$\mathfrak{J} =(-2.11,2.11)$. 
Suppose that
$Y_{2,k}\sim T_{2,k}(0,\cdot)$, then
$\mathbb{P}(Y_{2,k}\in \mathfrak{J}) \approx 0.43$, which is reduced
by almost a factor of two compared to the overlapping Gaussian
proposal. Further, if $\sigma = 0.25$ then
$\mathbb{P}(Y_{2,k}\in \mathfrak{J}) \approx 0.31$ and if
$\sigma = 0.05$ then
$\mathbb{P}(Y_{2,k}\in \mathfrak{J}) \approx 0.06$. Thus, the Plateau
proposals overlap less than the Gaussian proposals example and
further, the extent of the overlapping of the proposals is controlled
by the values of $\sigma$, $\sigma_1$, and $\sigma_2$.

Note that each Plateau proposal distribution has a support on
$\mathbb{R}$. This is to ensure that the support of the target
distribution is included within the support of the proposals. In
theory, this allows the Markov chain to explore the entire support of
the target distribution. In a practical setting, however, by selecting
the value of $\sigma$ appropriately, the tails of the distribution
decay to zero very quickly, making the inner proposals effectively
uniform distributions in view of numerical simulations. Therefore, in
practice one could elect to sample from these distributions using
direct draws from Uniform distributions to increase computation
speed. However, in all the experiments performed in this paper, we
sample from the exact Plateau proposal distributions defined above.

Note that the suggested Plateau proposals are very general; the choice
of proposal parameters, $\delta$, $\delta_1$, $\sigma$, $\sigma_0$ and
$\sigma_1$, allow for diverse sets of proposals. In the remainder of
the paper, we set $\sigma_0 = \sigma_1 = \varsigma$ for some value of
$\varsigma$, so that the decay of the tails of the outermost proposal,
$T_{M,k}$, are the same. Further, we fix the half-width of the
  proposals to be the same i.e.\ $\delta = \delta_1=\Upsilon $. Fixing
  these parameters means that the Plateau proposals are defined by 3
  parameters: $\Upsilon (=\delta = \delta_1)$, $\sigma$, and
  $\varsigma (=\sigma_0=\sigma_1)$.
The selected values for these parameters will determine the movement
of the Markov chain and thus its performance with respect to a
particular target distribution. In principle, one could set the values
of the parameters in an ad-hoc manner, e.g., a manually search over
the parameter space until a certain acceptance rate is achieved. A
practical approach is to automatically \textit{tune} the parameters as
the algorithm runs. In fact, in Section \ref{sec:adaptation-proposals}
an adaptation procedure that tunes the proposal parameter,
$\Upsilon (= \delta = \delta_1)$, is introduced. Thus only the 2
parameters $\sigma$, $\varsigma(=\sigma_0=\sigma_1)$ need to be
selected by the user. 
In simulations, we use $\sigma = 0.05$ and $\varsigma = 3$ so that
  there is minimal overlap of the Plateau proposals and to ensure the
  outermost proposal has heavy tails. Results using $\sigma = 0.05$
  and $\varsigma = 3$ resulted in good performance for a variety of
  target distributions.

\section{Adaptation of Plateau Proposals}\label{sec:adaptation-proposals}

As is typically the case when working with MCMC methods, the proposals
involving parameters need to be appropriately tuned for the algorithm
to be effective. 
Instead of manually tuning the parameters, an automated method can be
used to adapted the proposals as the MCMC procedure runs. These
adaptive methods use the information revealed so far to tune the
parameters of the proposals. For instance, \cite{haario2001adaptive}
proposes updating the covariance of a multivariate Normal proposal
using an empirical estimate of the covariance of the
target.
In the following, we discuss the adaptation mechanism for use with the
Plateau proposals introduced in Section \ref{sec:non-overl-prop}. The
adaptation mechanism is specifically chosen to exploit the
non-overlapping features of the Plateau proposals. In fact,
  combining the multiple-try paradigm with the localised shape of the
  proposals offers a natural adaptation criterion by monitoring
  preferred (component-wise) proposals, which will eventually allow to
  sample the state space in a more structured way.

In the innermost for-loop in Algorithm \ref{alg:very_generic} (steps 4
to 8) with Plateau proposals $T_{j,k}$, only one trial is selected
(step 6). The selected trial is associated with its
generating 
non-overlapping proposal. Over the Markov chain's iterations,
the frequency at which each 
trial is selected from a particular proposal can be monitored,
  which offers additional insight into the state space
  exploration. We advocate a procedure to update the Plateau
proposals to avoid two undesirable scenarios:
(i)
when the innermost proposal is selected
    too often; and
    (ii)
    when the outermost proposal is selected
    too often.
    Scenario~(i) suggests that the proposal distributions are too
    wide, such that trials are regularly being suggested near the
    previous state of the chain, i.e., the majority of moves are
    occurring in the interval $(x-\delta,x+\delta)$, when current state's component
    is $x_k=x$. Conversely, scenario~(ii) suggests that the proposal
    distributions are too narrow, such that the trials are regularly
    being suggested in the ``tails'', far away from the current position
    $x$.

    We suggest the following adaptation of the Plateau proposals
    $T_{j,k}$ to counteract these scenarios. As mentioned in
    Section~\ref{sec:non-overl-prop}, this adaptation procedure will
    change the half-width of all the Plateau proposals; namely the
    $\Upsilon=\delta =\delta_1$ parameter will be updated. First, adaptation can
    take place at regular, predefined intervals, of length $L$. Within these
    intervals each proposal is selected a number of times. Let
    $c_{j,k}^n$ denote the number of times $T_{j,k}$ was
    selected by the $n$th MCMC iteration.

    For scenario~(i), if $c_{1,k}^n > L \eta_1$ for some
    $\eta_1 \in (0,1)$, then the width of all the Plateaus is halved
    and the proposals are shifted closer to $x$ to leave no gaps. More
    precisely, the Plateau proposal parameters are updated as:
   $\Upsilon  \leftarrow 0.5\Upsilon$.

   For scenario~(ii), if $c_{M,k}^n > L \eta_2$ for some
   $\eta_2 \in (0,1)$, then the Plateaus widths are doubled and the
   proposals are shifted away from $x_k = x$ to leave no
   gaps. Formally, then the Plateau proposal parameters are updated
   as: $\Upsilon \leftarrow 2\Upsilon$.

    The proposed adaptation is summarised in Algorithm
    \ref{alg:adaptation} which can be inserted between steps 2 and 3
    of the MCMC Algorithm~\ref{alg:very_generic}. Note that the
    adaptation operation is performed every $L$ iterations.  At
    iteration $n$ the adaptation is performed with probability
    $P_n = \max\left\{0.99^{n-1},1/\sqrt{n}\right\}$. This ensures
    that the amount of adaptation reduces the longer the algorithm
    runs and thus satisfies the diminishing adaptation condition; see,
    e.g., \citep{roberts2007coupling}. Satisfying the diminishing
    adaptation condition ensures convergence of algorithm -- see
    Appendix \ref{sec:note-conv-adapt}.

  In summary, we advocate the use of the adaptation procedure,
    outlined in Algorithm \ref{alg:adaptation}, for our Plateau
    proposal MCMC. The adaptation procedure updates the
    $\delta = \delta_1$ parameter of the proposals. In simulations, we
    initialise these parameters at $ \delta = \delta_1 = 1$. As
    mentioned in Section \ref{sec:non-overl-prop}, the other proposal
    parameters are set at $\sigma = 0.05$ and $\varsigma = 3$.  
\begin{algorithm}[btp]
  \caption{Adaptation of MCMC}\label{alg:adaptation}
  \begin{algorithmic}[1]
    \REQUIRE{thresholds $\eta_1,\eta_2>0$; proposal parameter
      $\Upsilon = \delta=\delta_1$; iteration number $n$; adaptation
        interval length $L$} \IF{$n=1$} \STATE{Set
      $c_{1,k}^n = c_{M,k}^n = 0$ for all $k=1,\dots,d$}
    \ENDIF
    \STATE{Draw $r\sim\text{Uniform}(0,1)$}
     \IF{$\left[r <  P_n := \max(0.99^{n-1},1/\sqrt{n}) \right]$ and
       $\left[\left(n\mod L\right) = 0\right] $}
     \IF{$c_{1,k}^n > L \eta_1$}
     \STATE{Update: $\Upsilon \leftarrow 0.5\Upsilon$ 
}

     \ENDIF
     \IF{$c_{M,k}^n > L \eta_2$}
     \STATE{Update: $\Upsilon \leftarrow 2\Upsilon$
}

     \ENDIF
\STATE{Reset $c_{1,k}^n = c_{M,k}^n = 0$.}
\ENDIF
  \end{algorithmic}
\end{algorithm}

\section{Investigation of adaptation of MCMC methods}\label{sec:invest-adapt-mcmc}
Before assessing the long-term performance of the adaptive
  Plateau proposal MCMC in simulations, we first provide further
  insights into the effects of adaptation procedure. Specifically, we
  will investigate properties of the resulting Markov chain during the
  first, initial iterations to illustrate how these novel Plateau
  proposals adapt to a given target distribution.
For this investigation, two complementary
  scenarios are considered. First, we will study the resulting chain's coverage
  probability of a given target distribution's confidence region when
  the chain is initially started inside said region. Second, we will
  assess the first hitting time distribution for a chain to enter a
  high-probability region of the target distribution when initiated
  from a low-probability region.

  We will compare the adaptive Plateau proposal algorithm with the
  adaptive Gaussian MCMC
  algorithm~\cite{doi:10.1080/10618600.2018.1513365} and consider
  multivariate Normal target distributions.  As the target
  distribution is Normal, performance should favour the adaptive
  Gaussian MCMC method due to similarity between the target and
  proposal distribution shape. We stress that a particular adaptation
  strategy is proposal dependent, that is, here we compare the
  combined effect of 
  the proposal \textit{and} the adaptation procedure. We shall denote
  the adaptive Plateau proposal MCMC method as AP and
  the adaptive Gaussian proposal MCMC method proposed in
  \cite{doi:10.1080/10618600.2018.1513365} as AG2 (a variant of AG2,
  denoted as AG1, will be introduced later in Section
  \ref{sec:results}). Following
  \cite{doi:10.1080/10618600.2018.1513365}, the Gaussian proposal
  distributions in AG2 are adapted as follows: If the proposal with
  the largest standard deviation is under(over)-selected then it is
  halved (doubled). Conversely, if the proposal with the smallest
  standard deviation is under(over)-selected then it is doubled
  (halved). After either of these two updates, the other standard
  deviations are adjusted to be equidistant on a log-scale (base
  2). For further algorithmic details on this adaptation scheme; see
  \cite{doi:10.1080/10618600.2018.1513365}. The adaptation interval
  for both methods was set to $L=50$ and $P_n = -1$ (in Step 4, Algo
  2) to trigger adaptations every $50$ iterations. These
  settings were selected in order to make a fair comparison between
  the two MCMC methods.  No burn-in was used in these simulations as
  we are investigating the performance of the MCMC algorithms during
  the initial iterations.

\begin{figure}[btp]
  \centering
   \includegraphics[width=0.6\linewidth]{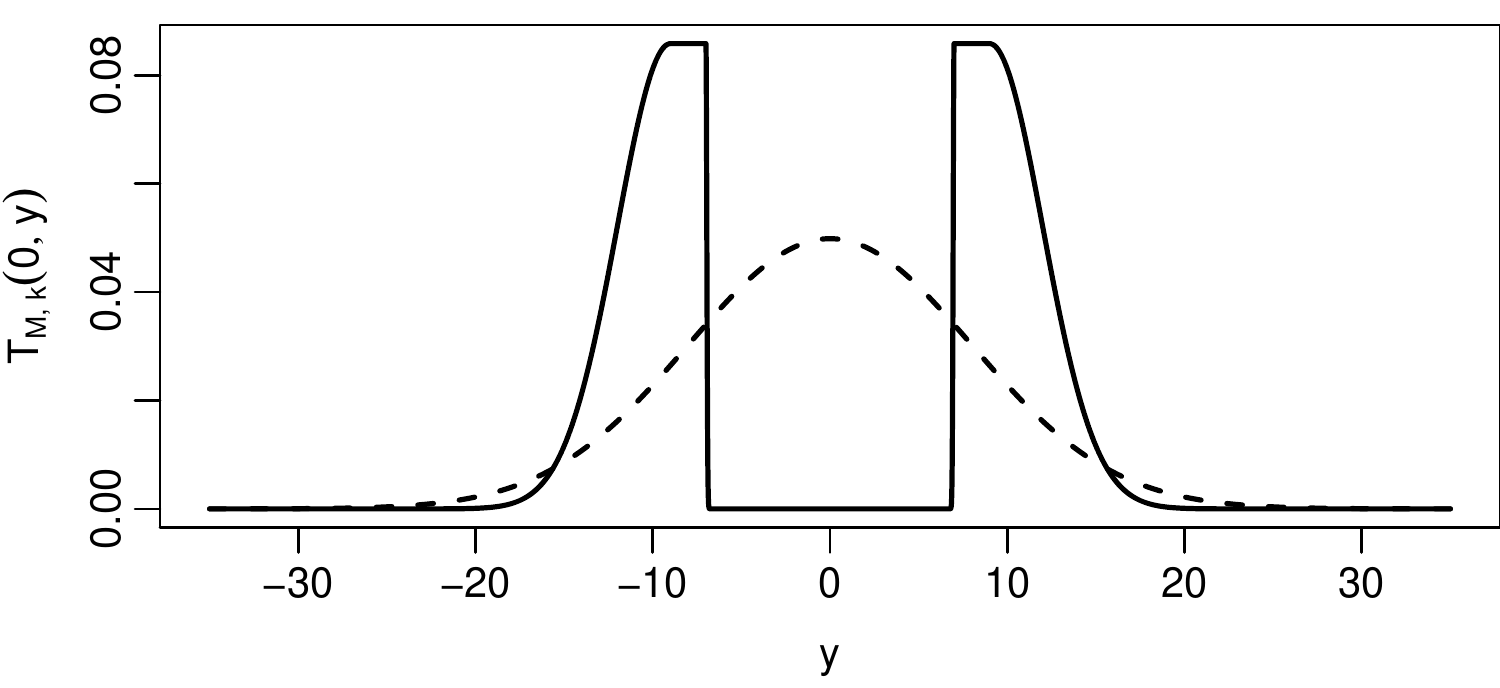}

   \caption{Outermost proposal density ($j=M=5$) for the AP method (solid) and
     the AG2 method (dashed) upon initialisation of the MCMC
     algorithms.} \label{fig:presim_illu}
\end{figure}

  For the simulation outcomes that follow we fix the number of trials
  $M=5$. The proposals' standard deviations in the AG2 method are
  initialised with $2^{j-2}$ for proposal $j=1,\dots,M$. We
    reiterate that the adaptively chosen Plateau parameters are
  initialised with $\delta = \delta_1=1$, while the other
    Plateau parameters are set to $\sigma = 0.05$ and
  $\varsigma = 3$. Moreover, we use the thresholds
  $\eta_1 = \eta_2 = 0.4$ for the AP method. The outermost proposal
  for both the AP and AG2 methods of this particular
    initialisation are illustrated in
  Fig.~\ref{fig:presim_illu}. Notice that the Gaussian proposal has
  slightly heavier tails that may give the AG2 an advantage by
    allowing larger moves compared to the AP method.

\subsection{Scaling of the Proposals and Coverage Probability}
As mentioned above, we first consider the MCMC methods'
  coverage probabilities by monitoring how frequently a
  high-probability region is visited. We begin with a target
  distribution, $\varpi_1$, given by a 5-dimensional normal
  distribution with mean zero and covariance matrix
  $\Sigma = \text{diag}(0.001,0.1,1,10,100)$. Note that the different
  magnitudes of the variances of the target will be
  captured directly by both MCMC methods. This is because the
  component-wise updates of the algorithms are along the same components
  of the target distribution; namely on $x_1,x_2,\dots,x_5$.  

\begin{figure}[tbp]
  \centering

 \subfloat[Component $k=1$]{
   \includegraphics[width=0.49\linewidth]{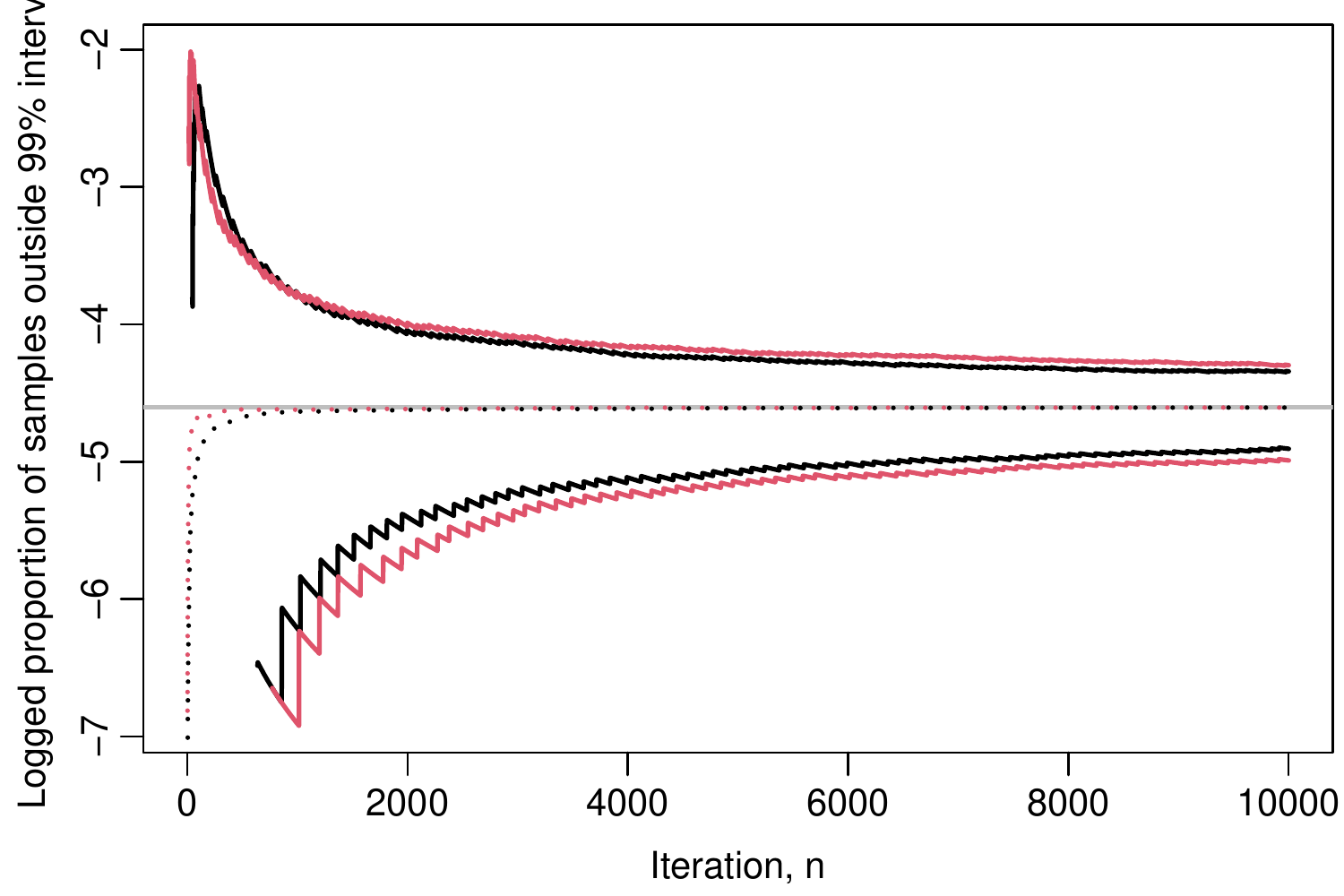}\label{fig:presim2-1}
 }
  \subfloat[Component $k=2$]{
 \includegraphics[width=0.49\linewidth]{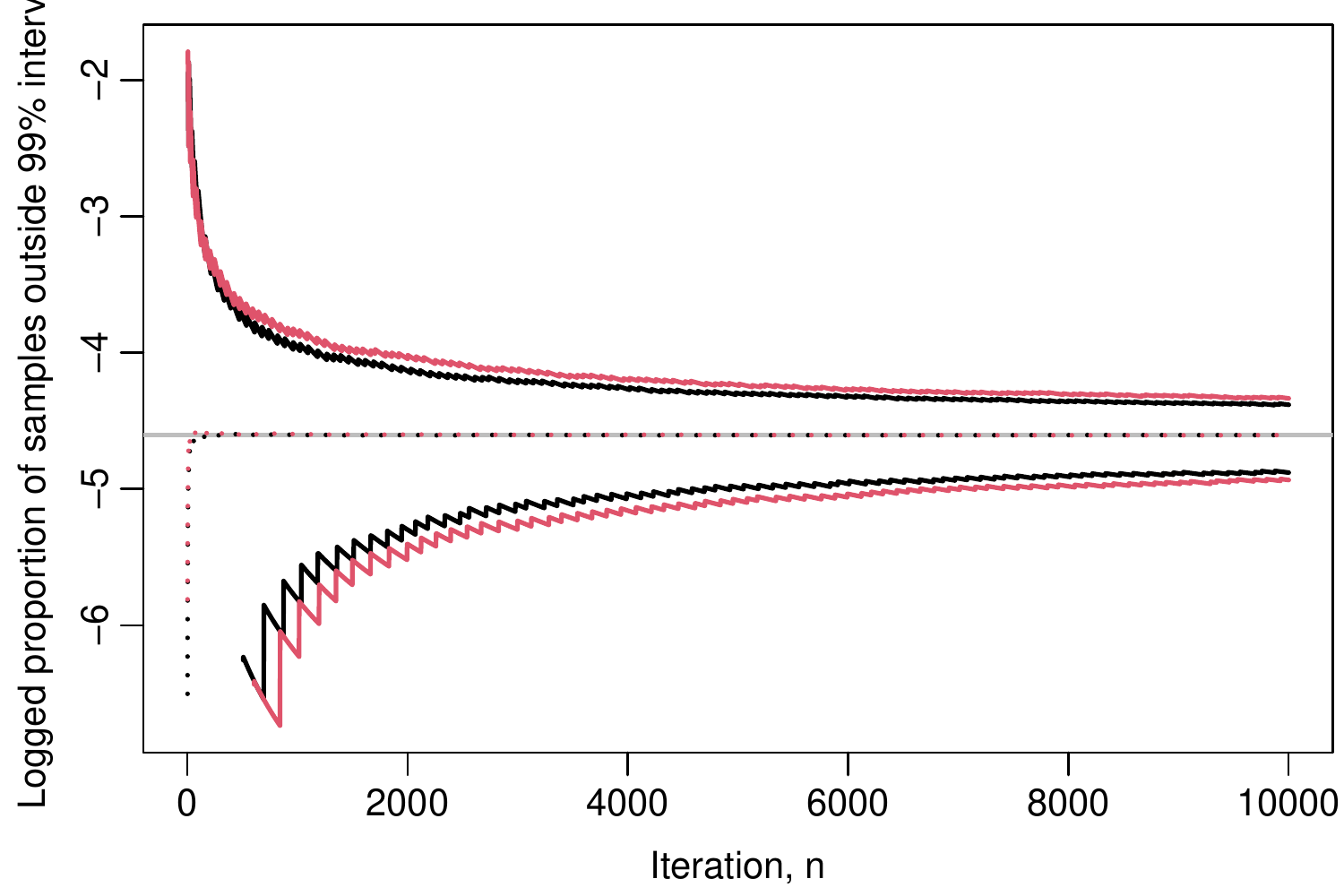}\label{fig:presim2-2}
}

 \subfloat[Component $k=3$]{
   \includegraphics[width=0.49\linewidth]{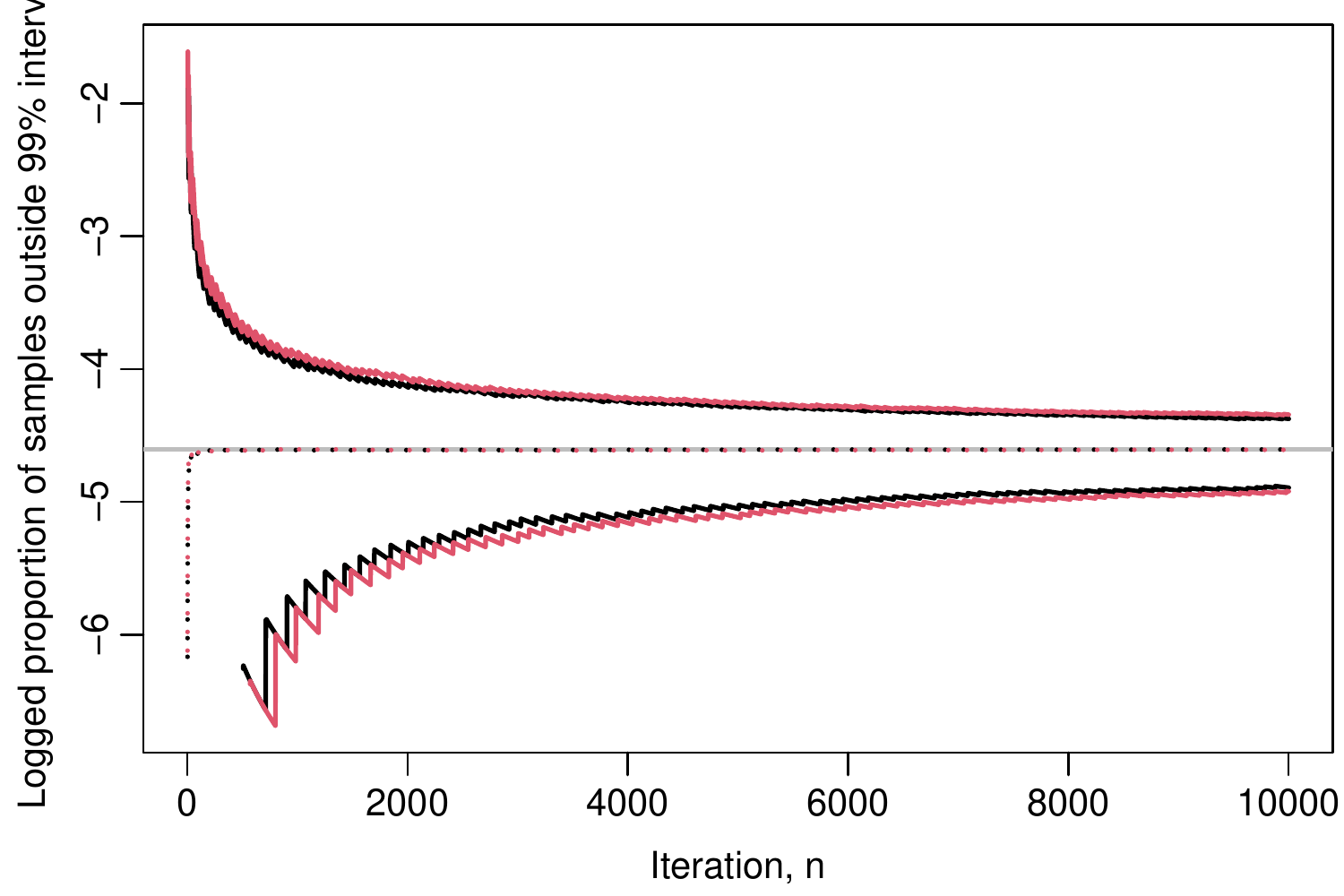}\label{fig:presim2-3}
}
 \subfloat[Component $k=4$]{
 \includegraphics[width=0.49\linewidth]{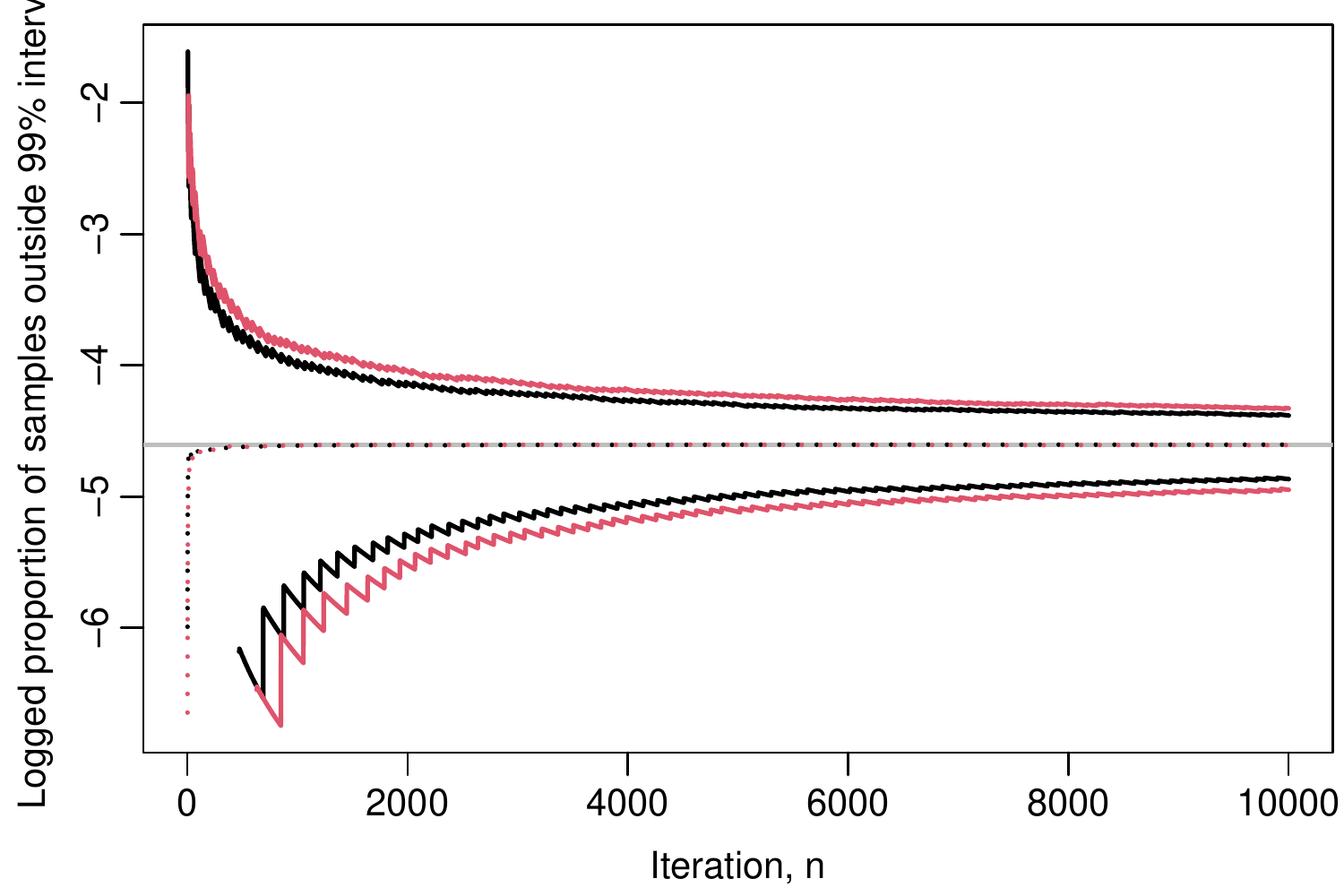}\label{fig:presim2-4}
}

 \subfloat[Component $k=5$]{
   \includegraphics[width=0.49\linewidth]{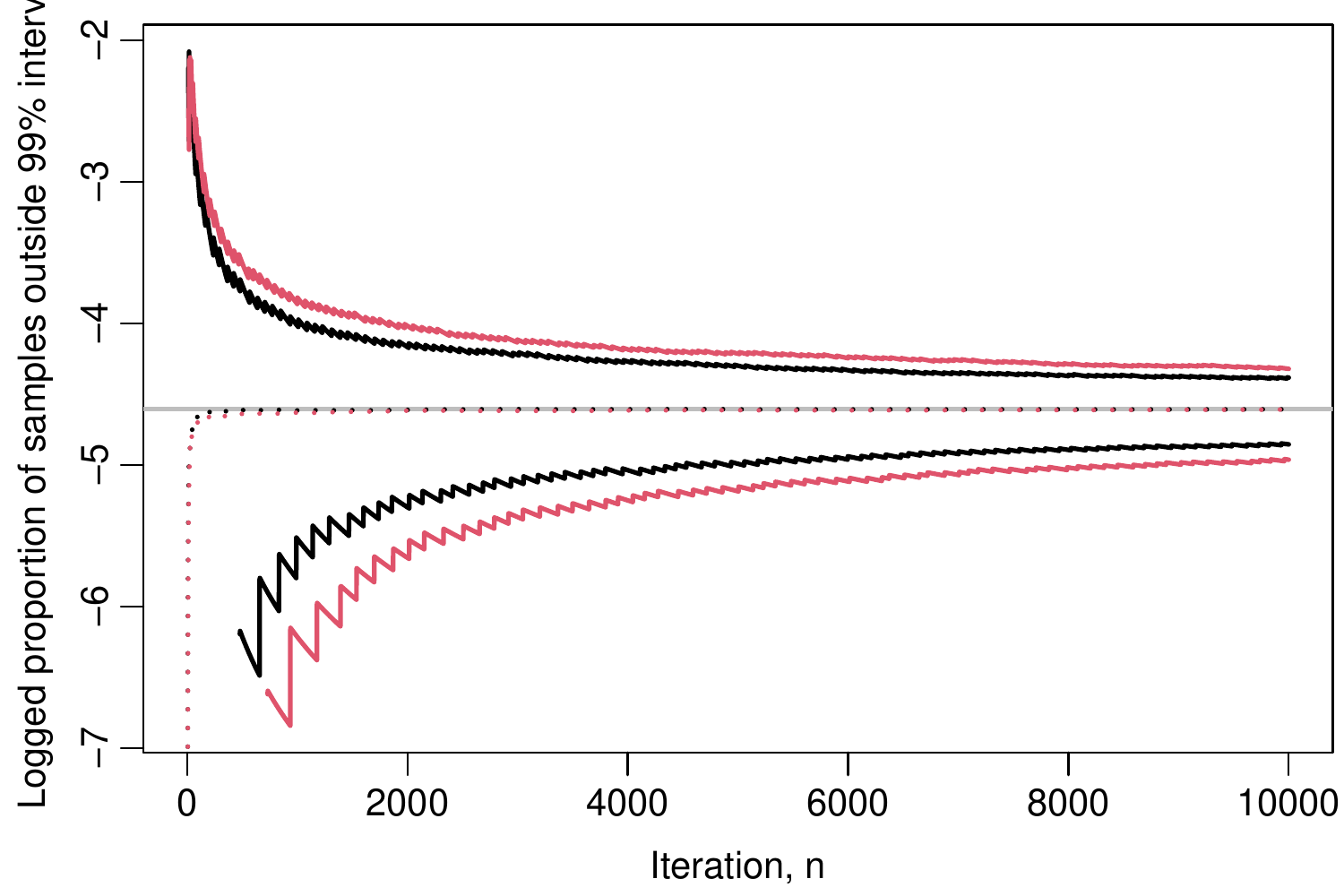}\label{fig:presim2-5}
 }
  \subfloat[Joint]{
\includegraphics[width=0.49\linewidth]{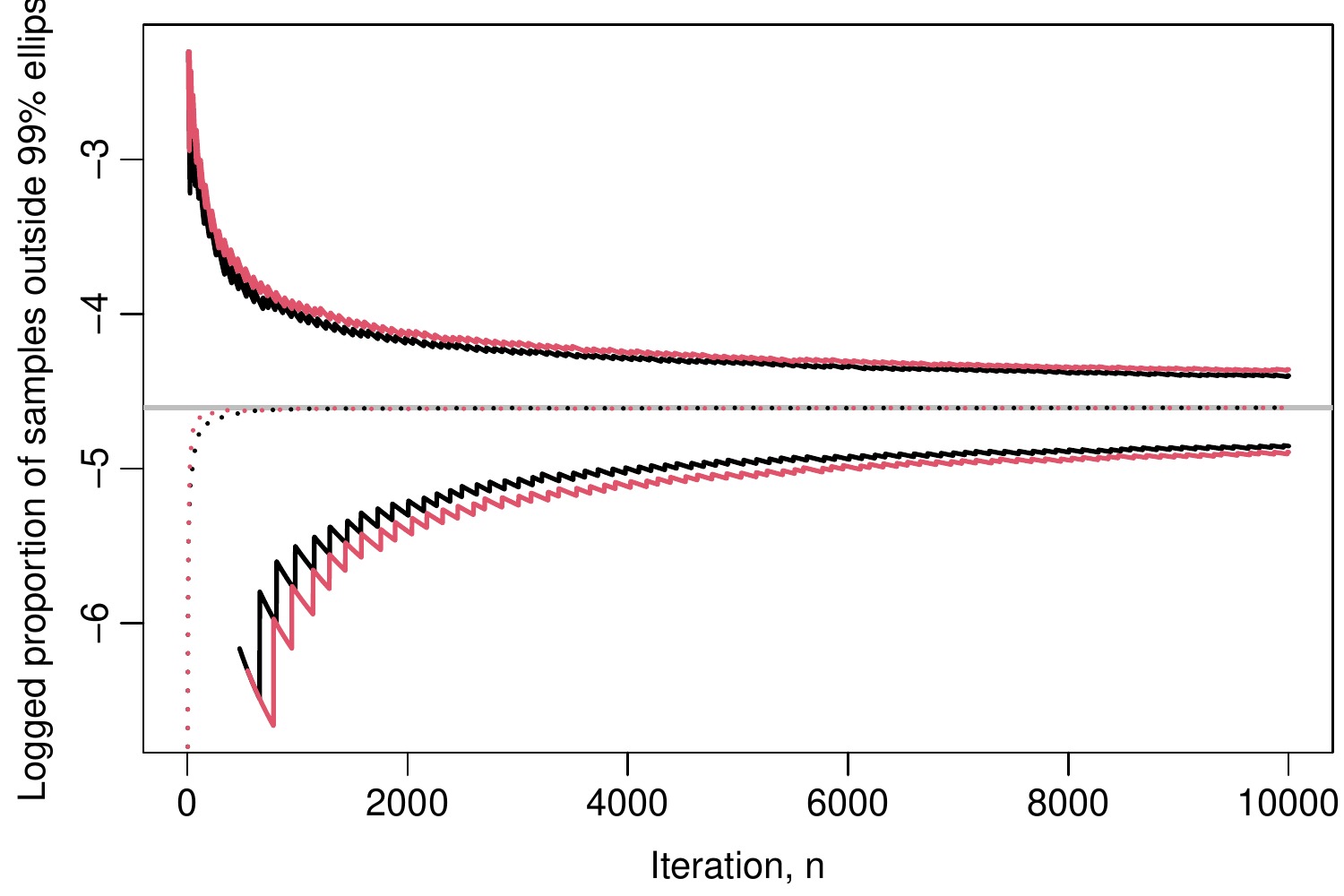}\label{fig:presim2-all}
}
 
  \caption{\textbf{Uncorrelated 5-dimensional Gaussian Target, $\varpi_1$:} Figures (a) to (e) presents the average and 95\% empirical quantiles of $\{C_{n}^{(r)}(k);
    r=1,\dots,R \}$ for $k=1,\dots,5$ respectively. Figure (f) present
    the average and 95\% empirical quantiles of $\{D_{n}^{(r)};
r=1,\dots,R \}$. Averages are represented by dotted lines and
quantiles by solid lines. All results are presented on the log-scale. Black represents the AP method and red
represents the AG2 method.} \label{fig:pre_sim_2}
\end{figure}

The number of independent repetitions of the MCMC simulations
was set to $R=5,000$ and the number of MCMC iterations performed for each
run was $N=10,000$.  For each MCMC method denote the chain for the
$r$th iteration as
$\bs X_{0}^{(r)}, \bs X_{1}^{(r)},\dots,\bs \dots,\bs X_{N}^{(r)}$
where
$\bs X_{j}^{(r)} = \left( X_{j,1}^{(r)},\dots, X_{j,5}^{(r)}\right)^T$
for $r= 0,\dots,R$. Further, denote the component-wise variances of
the target $\varpi_1$ as $\sigma_1^2,\dots,\sigma_5^2$. Each
repetition $r$ of the chain was started at the origin
$\bs X_{0}^{(r)} = \bs X_0 = (0,0,0,0,0)^T$ for both methods. For each
repetition $r$, we report summaries of the running
  ($n=1,\dots, N$) empirical coverage probabilities of the
  component-wise ($k=1,\dots,5$) confidence regions
\begin{equation*}
  C_n^{(r)}(k) := \frac{1}{n}\sum_{j=0}^n \mathrm{I}\left(\frac{(X_{j,k}^{(r)})^2}{\sigma^2_k}
  > z_1\right),
\end{equation*}
where $z_1$ is such that $P(Z_1>z_1)= 0.99$ and $Z_1\sim \chi^2_{1}$
as well as the running empirical coverage probability of the
  joint 99\% confidence region defined as
\begin{equation*}
  D_n^{(r)} := \frac{1}{n}\sum_{j=0}^n \mathrm{I}\left( (\bs X_{j}^{(r)})^T\Sigma^{-1}\bs X_{j}^{(r)}
  > z_2\right),
\end{equation*}
where $z_2$ is such that $P(Z_2>z_2)= 0.99$ and $Z_1\sim \chi^2_{5}$.
More precisely, for each $n=1,\dots,N$ the average and the 2.5\% and
97.5\% empirical quantiles of $\{C_{n}^{(r)}(k); r=1,\dots,R \}$ for
$k=1,\dots,5$ and of $\{D_{n}^{(r)}; r=1,\dots,R \}$ are reported in
Fig.~\ref{fig:pre_sim_2}. Note that the vertical axes is on the
log-scale to improve visibility of the results. In each reported
summary, a point-wise interval is created by the 2.5\% and 97.5\%
quantiles. The resulting empirical confidence intervals for
the AP method are, for the majority of iterations, narrower than those
for the AG2 method. This indicates that the AP method is better than
the AG2 method in terms of the coverage of the 99\% marginal
distributions of the target as well as the entire joint
distribution. We note that the AG2 method uses Gaussian component-wise
proposals share exactly the same shape as the marginals of the target
distribution. Despite this advantage, the AP method produces better
results in terms of coverage, as presented here.

\begin{figure}[tbp]
  \centering

 \subfloat[Component $k=1$]{
   \includegraphics[width=0.49\linewidth]{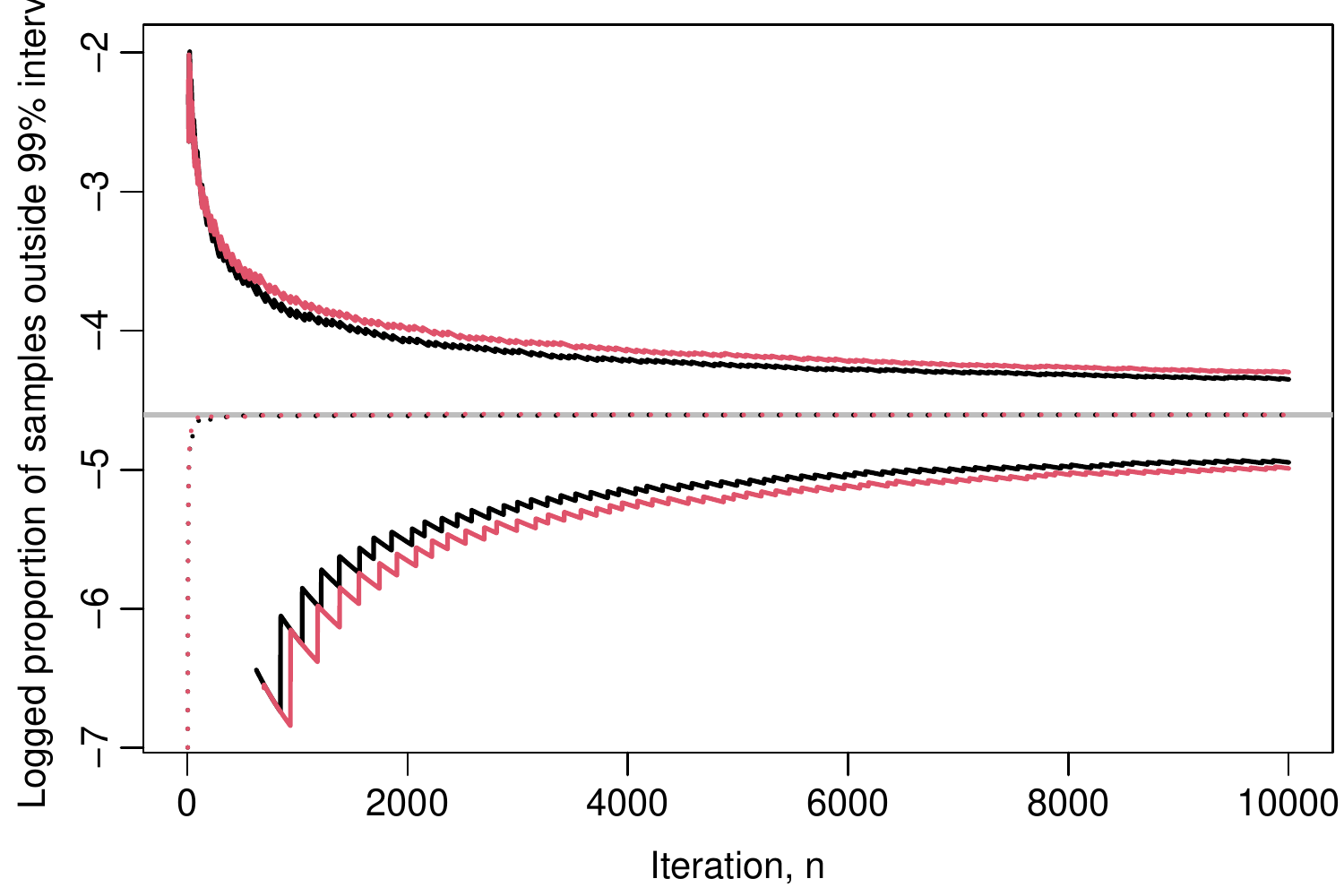}\label{fig:presim2-1}
 }
  \subfloat[Component $k=2$]{
 \includegraphics[width=0.49\linewidth]{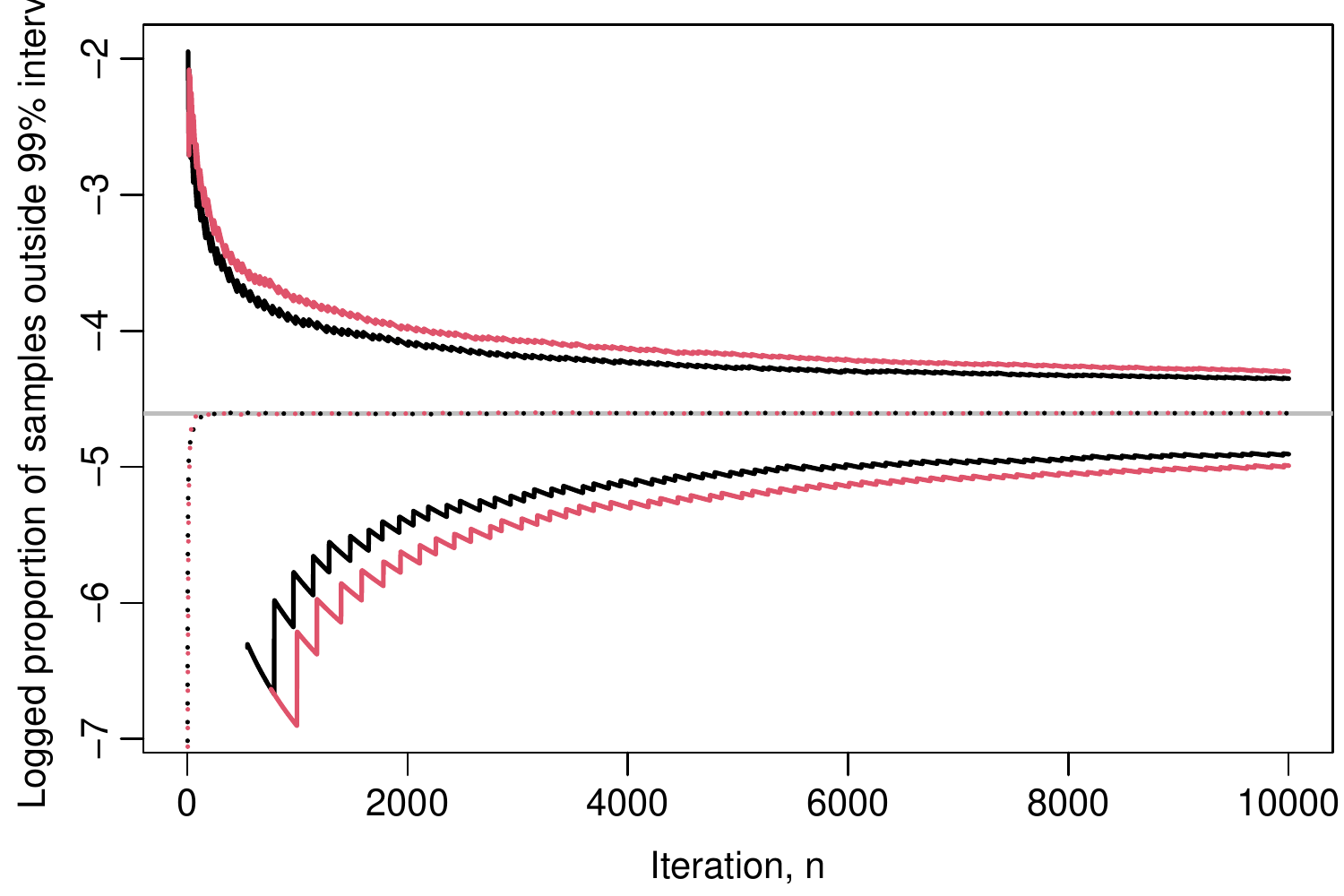}\label{fig:presim2-2}
}

 \subfloat[Joint]{
   \includegraphics[width=0.49\linewidth]{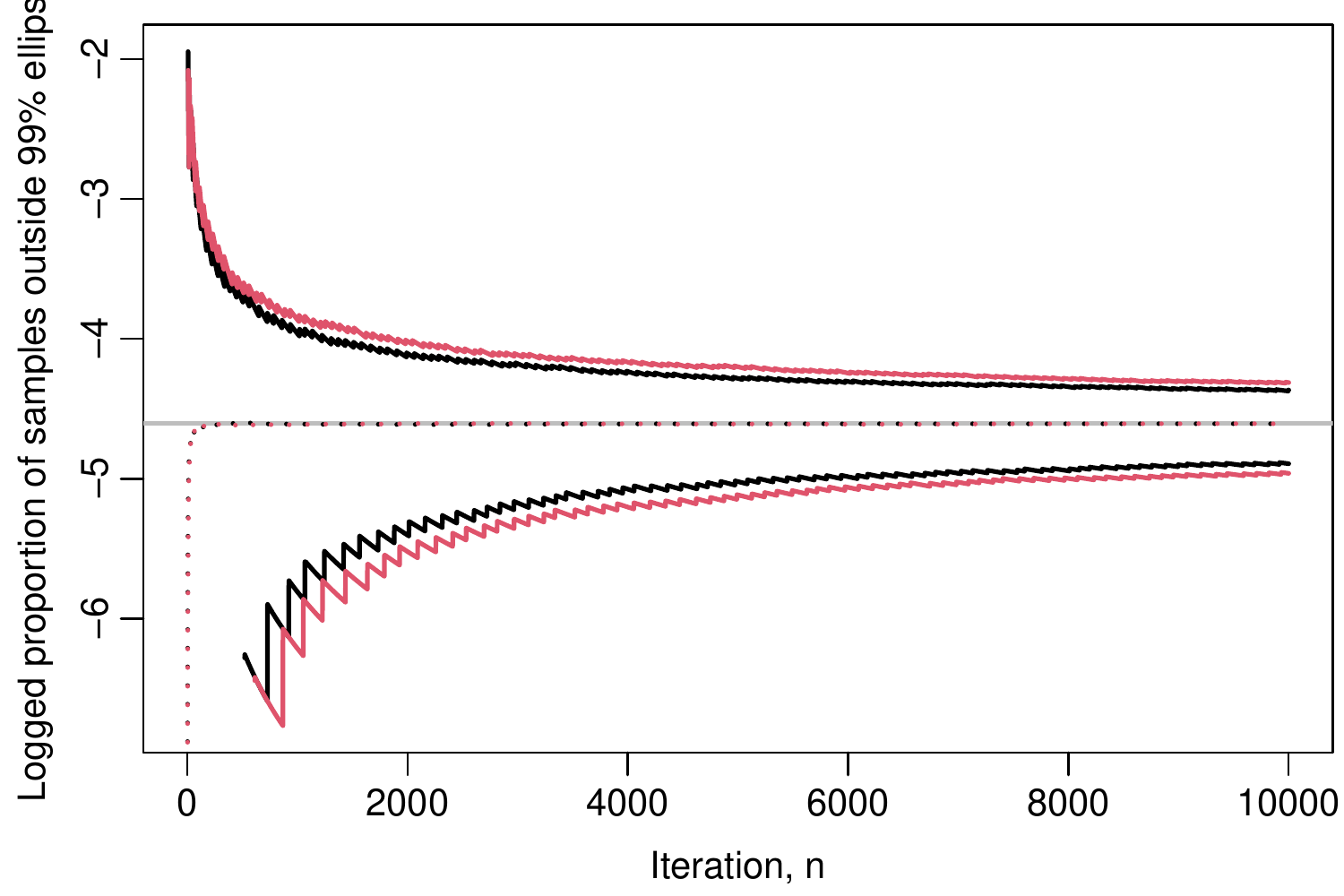}\label{fig:presim2-3}
}
 
  \caption{\textbf{Correlated 2-dimensional Gaussian Target,
      $\varpi_2$:} Figures (a) and (b) presents the average and 95\% empirical quantiles of $\{C_{n}^{(r)}(k);
    r=1,\dots,R \}$ for $k=1,2$ respectively. Figure (c) presents
    the average and 95\% empirical quantiles of $\{D_{n}^{(r)};
r=1,\dots,R \}$. Averages are represented by dotted lines and
quantiles by solid lines. All results are presented on the log-scale. Black represents the AP method and red
represents the AG2 method.} \label{fig:pre_sim_3}
\end{figure}

Next, we explore the performance of the adaptive MCMC
algorithms for a target distribution, $\varpi_2$, with correlation
across the components. The target distribution for this
  second example is taken to be a bivariate Normal distribution with
  mean zero and covariance matrix
\begin{equation*}
  \Sigma = \left(
    \begin{array}{cc}
      0.25&1.875\\
      1.875&25
    \end{array}
\right).
\end{equation*}
Note that in this simulation, the component-wise MCMC proposal
updates, for both the AP and AG2 method, are not aligned with the
principal components of the target distribution. Therefore, this
target distribution should pose a greater difficulty to sample from in
comparison to the previous target, $\varpi_1$. The results from this
simulation study are presented in Fig.~\ref{fig:pre_sim_3}, where we
report the analogous summaries of the running empirical coverage
probabilities, as for the previous example. Recall that both MCMC
methods adapt independently per component. Therefore, the correlation
between the components is ignored in the MCMC methods and their
adaptations.  Despite ignoring the correlation, both MCMC methods
perform well in terms of coverage probabilities.  As shown in
Fig.~\ref{fig:pre_sim_3}, the AP method consistently outperforms the
AG2 method, in terms of the width of the empirical confidence
intervals.

\subsection{Moving to a High Density Region of Target}

We now investigate the adaptive MCMC methods' ability to move
toward areas of high target density, when initially started in
  at a state outside that region. For this study, the target
distribution is again taken as the correlated bivariate
normal distribution, $\varpi_2$. The number of independent
  repetitions of the MCMC runs was again set to $R=5,000$ and the
number of MCMC iterations performed each run was $N=1,000$. Each run was
initialised at the location very far away from the mean,
$\bs X_0 = (50,50)^T$.

\begin{figure}[btp]
  \centering
  \includegraphics[width=0.7\linewidth]{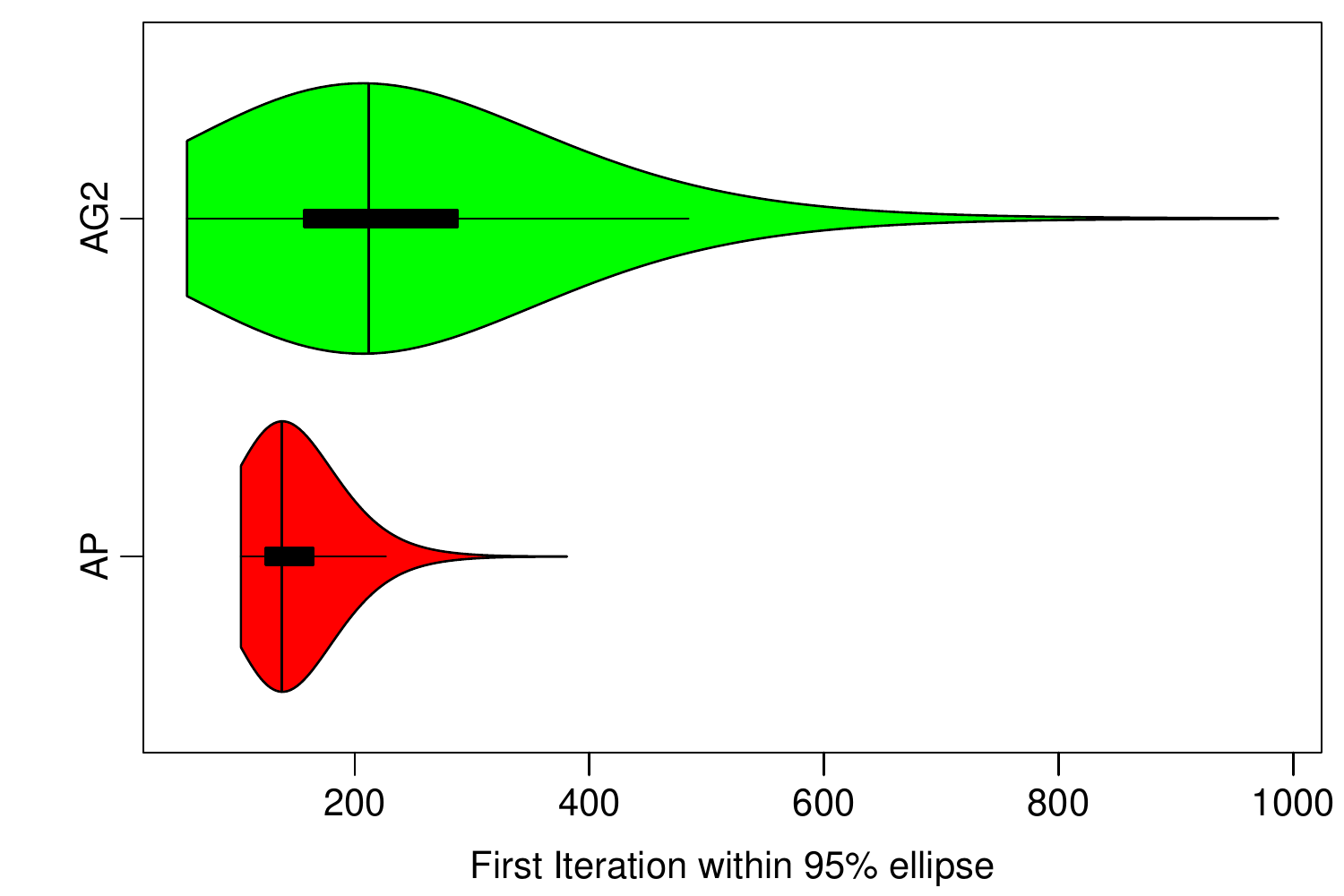}
  \caption{\textbf{Correlated 2-dimensional Gaussian Target,
      $\varpi_2$:} Violin plot of first hitting time of the Markov
    chain into 95\% ellipse. AP results represented in red and AG2 results in blue.}
  \label{fig:pre_sim_1}
\end{figure}

For each repetition $r$ of the MCMC method, the first time
the Markov chain entered the (joint) 95\% ellipse of the target
is recorded as
\begin{equation*}
  J^{(r)} := \min\left(j\in\{0,\dots,N\}; (\bs X_{j}^{(r)})^T \Sigma^{-1}  \bs X_{j}^{(r)}  <   z_0 \right),
\end{equation*}
where $z_0$ is such that $P(Z_0>z_0)= 0.95$ and $Z_0\sim
\chi^2_2$. The empirical distribution of the first hitting times
$\{J^{(1)},\dots,J^{(R)}\}$ is summarised in the violin plots
presented in Fig.~\ref{fig:pre_sim_1} for AP and AG2 methods. Violin
plots \citep{doi:10.1080/00031305.1998.10480559} are boxplots with
kernel density estimates attached to the sides. We use violin plots as
opposed to just boxplots to give a better illustration of the shape of
the distributions. In the violin plots the median is represented by a
vertical line.
From Fig.~\ref{fig:pre_sim_1} we observe that the AP method
consistently moves into the 95\% ellipse earlier than the AG2
method. Although the AG2 method does enter the ellipse earlier in some
iterations, the AP method always ($5,000$ out of $5,000$ iterations) entered
the high-density region in less than $381$ iterations. In comparison, in
$517$ out of $5,000$ iterations, the AG2 took more than $381$ iterations to
enter the ellipse.

\section{Results}\label{sec:results}
In this section, we compare the long-term performance of the proposed
adaptive component-wise, multiple-try MCMC method with other MCMC
methods in simulations. The adaptive Plateau MCMC (AP) is compared a
Metropolis-Hastings algorithm with Gaussian proposals (MH) and two
versions of an adaptive Gaussian MCMC (AG1 and AG2)\, as introduced in
\cite{doi:10.1080/10618600.2018.1513365}. The difference between AG1
and AG2 is that AG1 uses \eqref{eq:lambda} with $\alpha = 2.5$ (i.e.,
the same weight as in AP), while AG2 uses $\alpha = 2.9$ as is
suggested in \cite{doi:10.1080/10618600.2018.1513365}. The proposal
distributions in both AG1 and AG2 are adapted in the fashion outlined
in Section \ref{sec:invest-adapt-mcmc}.

For all simulations and methods with multiple trials we fix the number
of trials $M=5$. Investigation of the methods performance for
differing values of $M$ is beyond the scope of this paper,
which we leave for future work. However, interesting
discussions in that direction already exist, see \cite{Martino2017}
for example.

The proposal standard deviation in the AG1 and AG2 methods are initialised at $2^{j-2}$ for $j=1,\dots,M$. The
Plateau parameters values are initialised at $\delta = \delta_1=1$,
$\sigma = 0.05$ and $\varsigma = 3$ with $\eta_1 = \eta_2 = 0.4$ for
the AP method.

The proposal distributions used in the MH method depend on the
particular target distribution and the choices used are
  summarised in Table~\ref{tab:MH-proposals} below. The various
target distributions considered in the simulations are now introduced.

\subsection{Target Distributions}
In order to compare the aforementioned methods, we investigate their
performances by applying them to sample from a variety of target
distributions.

\subsubsection*{Mixture of Gaussians}
Consider a mixture of two 4-dimensional Gaussians
\begin{equation*}
  \frac{1}{2}\normal(\bs\mu_1,\Sigma_1)+  \frac{1}{2}\normal(\bs\mu_2,\Sigma_2),
\end{equation*}
where
\begin{equation*}
  \bs\mu_1 = (5,5,0,0)^T,\quad   \bs\mu_2 = (15,15,0,0)^T
\end{equation*}
and
\begin{equation*}
  \Sigma_1 = \text{diag}(6.25,6.25,6.25,0.01),\quad   \Sigma_2 = \text{diag}(6.25,6.25,0.25,0.01).
\end{equation*}
We refer to this target distribution as $\pi_1$.

\subsubsection*{Banana Distribution}
Consider the 8-dimensional ``banana-shaped'' distribution
\citep{Haario1999}, which is defined as follows. Let $f$ be the density
of the $8$D normal distribution $\normal(\bs 0,\Sigma_3)$ with
covariance given by $\Sigma_3 = \text{diag}(100,1,\dots,1)$. The
density function of the banana distribution with non-linearity
parameter $b>0$ is given by
$  f_b=f \circ \phi_b$
where the function $\phi_b$ is
\begin{equation*}
  \phi_b(\bs x) = (x_1,x_2+b x_1^2 - 100
  b,x_3,\dots,x_8)\quad\text{for }\bs x\in\mathbb{R}^8.
\end{equation*}
The value of $b$ determines the amount of non-linearity of $\phi_b$.  Here, we consider the target distribution $\pi_2 = f_{0.03}$. Thus,
the first 2 components of this distributions are highly correlated --
see Figure \ref{fig:a}.

\subsubsection*{Distributions perturbed by oscillations}

\begin{figure}[H]
  \centering
 \subfloat[][Joint PDF of components $1$ and $2$ of $\pi_1$.]{\includegraphics[width=0.49\textwidth]{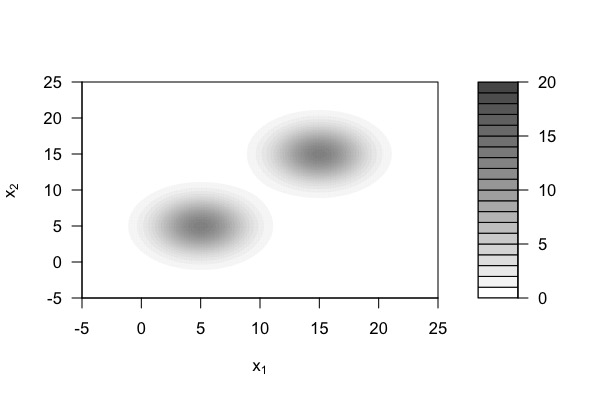}\label{fig:a}}
 \subfloat[][Joint PDF of components $1$ and $2$ of $\pi_2$.]{\includegraphics[width=0.49\textwidth]{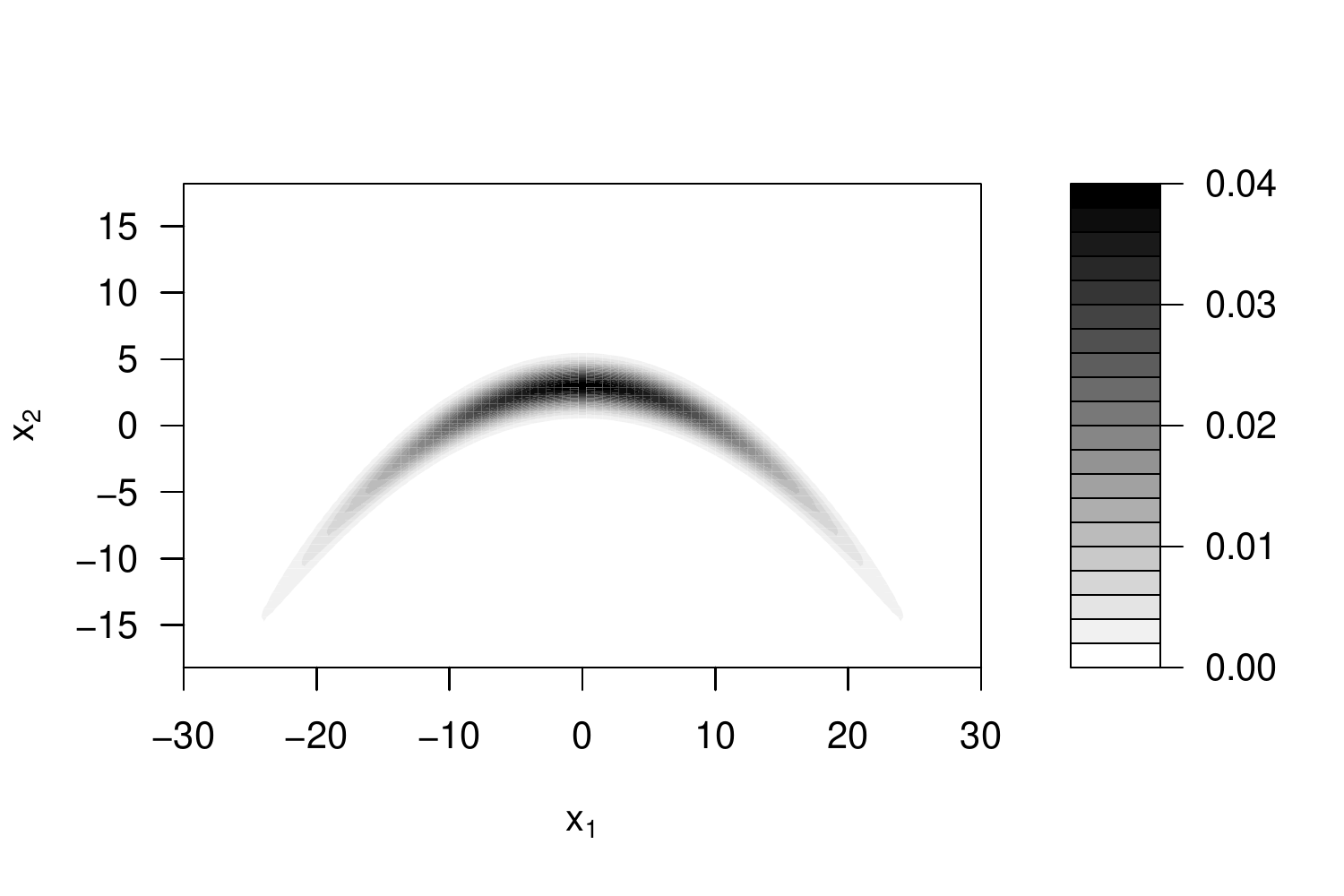}\label{fig:b}}

 \subfloat[][Un-normalised PDF of $\pi_3$.]{\includegraphics[width=0.49\textwidth]{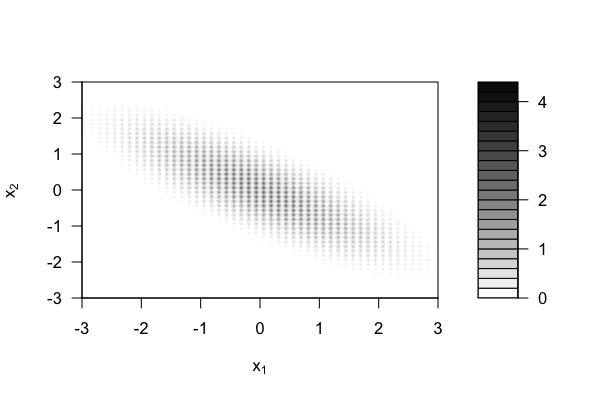}\label{fig:c}}
 \subfloat[][PDF of $\pi_4$.]{\includegraphics[width=0.49\textwidth]{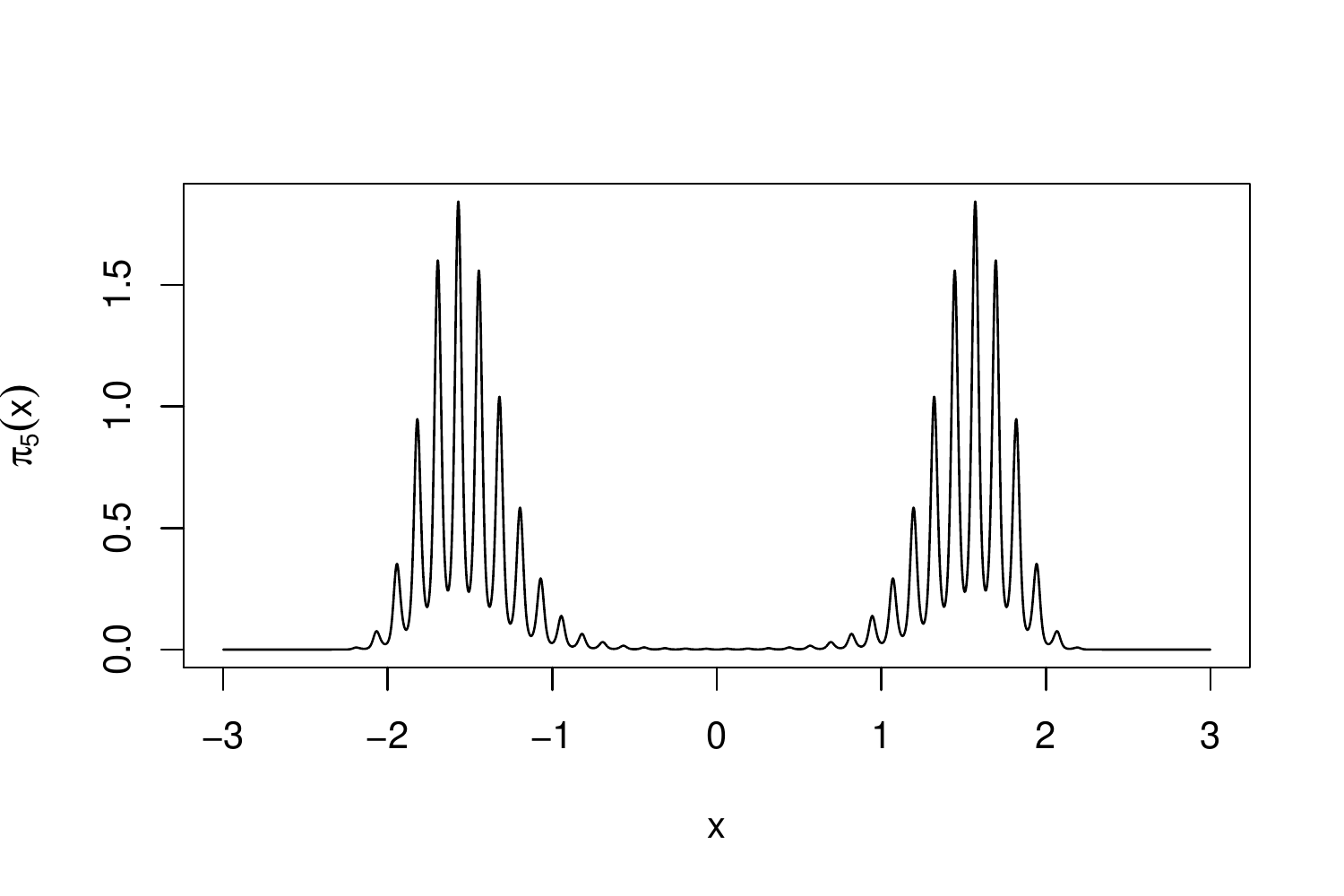}\label{fig:d}}
  \caption{Selected marginal density plots of simulation target distributions}
  \label{fig:targets}
\end{figure}

Another target we consider is the perturbed 2-dimensional Gaussian, whose
probability density function is given by
\begin{equation*}
  \pi_3(\bs x)\propto \exp\left[-\bs x^TA \bs x -
    \cos\left(\frac{x_1}{0.1}\right) -
    0.5\cos\left(\frac{x_2}{0.1}\right) \right] =:
  \widetilde{\pi}_3(\bs x)\quad\text{for }\bs x\in\mathbb{R}^2
\end{equation*}
where
\begin{equation*}
  A=\left(
    \begin{array}{cc}
      1&1\\
      1&3/2
    \end{array}
\right).
\end{equation*}
Figure \ref{fig:c} displays the un-normalised function
$\widetilde{\pi}_3(\bs x)$. Lastly, we also consider the following
perturbed version of the 1D bi-stable distribution
$x\mapsto Z^{-1} e^{-x^2+5x^2}$, whose PDF is given by
\begin{equation*}
  \pi_4(x)\propto \exp\left[-x^4 + 5x^2  - \cos\left(\frac{x}{ 0.02}\right) \right]\quad\text{for }x\in\mathbb{R}.
\end{equation*}
Figure \ref{fig:d} displays the PDF of $\pi_4(x)$ where the
normalising constant is approximated by numerical integration.

\subsection{Run parameters}
Each simulation run of the MCMC methods was independently repeated
$R=200$ times and for each run a burn-in period of $50$\% of the MCMC
iterations was used. During the burn-in period, the AP, AG1 and AG2
were allowed to adapt their proposals. For each repetition, all
methods started at the same random initial condition $\bs{x}_0$. The
number of MCMC iterations, $N$, used for each method is presented in Table
\ref{tab:sim_N}, which was determined by a trial run. In order to make
fair comparisons, the number of MCMC iterations performed by
Metropolis-Hastings algorithm is $d\times M$ times larger than the
multiple-try versions. This is because multiple-try methods cycle over
all $d$ components and evaluates the target for $M$ trials each
iteration. This will ensure that the number of times the target
distribution is evaluated by each MCMC method is exactly same and that the
computational effort is approximately the same.
\begin{table}[H]
  \centering
  \begin{tabular}{|c|c|c|c|}\hline
    Target & Dimensions ($d$) & Adaptive & MH\\\hline\hline
    $\pi_1$& 4 &4,000   & 80,000 \\
    $\pi_2$& 8 &10,000   & 400,000\\
    $\pi_3$& 2 &3,000   & 30,000   \\
    $\pi_4$& 1 &3,000    & 15,000 \\\hline
  \end{tabular}
  \caption{Number of MCMC iterations used in simulations for each target distribution.}\label{tab:sim_N}
\end{table}
The proposal distribution in the MH method for each target is
presented in Table \ref{tab:MH-proposals}. Note that these proposals
are based on the target distribution, which would be typically be
unknown in practice. Consequently, the MH method can be viewed as
being optimally tuned.

The particular scaling of $2.4/\sqrt{d}$ follows from
\cite{gelman1996efficient}.

\begin{table}[H]
  \centering
  \begin{tabular}{c|c}
    Target & Proposal Distribution \\\hline\\
    $\pi_1$ & $\displaystyle \frac{2.4}{\sqrt{4}}\left[ 0.5
              \normal(\bs 0,\Sigma_1) + 0.5\normal(\bs 0,\Sigma_2) \right]$\\\\\hline\\
    $\pi_2$ & $\displaystyle  \frac{2.4}{\sqrt{8}} \normal(\bs 0,\Sigma_3)$\\\\\hline\\
    $\pi_3$ & $ \displaystyle  \frac{2.4}{\sqrt{2}} \normal(\bs 0,A^{-1})$\\\\\hline\\
    $\pi_4$ & $ \displaystyle  2.4 \normal(0,1)$\\\\\hline
  \end{tabular}
\caption{Proposal distributions used in the Metropolis-Hastings algorithm.}\label{tab:MH-proposals}
\end{table}

\subsection{Simulation Results}

For each target distribution, we compare the performance of the MCMC
methods using measures which we now define. Denote the Markov chain
produced by one of the MCMC methods for the $r$th independent
repetition as $\bs X_0^{(r)},\dots,\bs X_N^{(r)}$ where
$\bs X_i^{(r)} = (X_{i,1}^{(r)},\dots, X_{i,d}^{(r)})^T$ for
$r=1,\dots, R$. Denote  the component-wise variances of the target
$\pi$ as $\sigma_1^2,\dots,\sigma_d^2$.

We will use the integrated autocorrelation time (ACT) of the MCMC
methods as a measure of performance. The ACT for the chain's
  $k$th component is given by
  \begin{equation*}
  \mathrm{ACT}_{k} = 1 + \frac{2}{\sigma^2_k}\sum_{i=1}^N \cov(
  X_{0,k},  X_{i,k})\;,
\end{equation*}
provided that the chain is stationary so that $\bs{X}_0\sim \pi$. For
every repetition $r=1,\dots, R$ of the MCMC method, the integrated
autocorrelation times are estimated based on the observed Markov chain
$\bs X_0^{(r)},\dots,\bs X_N^{(r)}$ component-wise using the initial
sequence estimator introduced in \cite{practical_MCMC_geyer}. With
slight abuse of notation, we will denote the resulting chain-based
autocorrelation times by $ \mathrm{ACT}_{k}^{(r)}$. Smaller
autocorrelation times indicate that consecutive samples have lower
correlation. Autocorrelation times are inversely proportional to the
effective sample size \cite{kong1992note,kong1994}, which is commonly
used as a measure of performance. In fact, the effective sample size
is often interpreted as the number of samples that would need to be
(directly) drawn from the target in order to achieve the same variance
as that from an estimator of interest using independent
samples. Higher effective sample sizes and therefore lower
autocorrelation times are desirable. Another way of interpreting the
ACTs is through the accuracy of a chain-based Monte Carlo
integration. As a matter of fact, the mean squared error of a Monte
Carlo estimator can be expressed as a sum of the component-wise ACTs
weighted by the component-wise variance. Consequently, a method with
lower ACTs will offer more accurate Monte Carlo integration for the
same chain length. It is moreover noteworthy that an MCMC method's
ACTs also characterise the asymptotic variance (so-called
time-averaged variance constant in this context) of an Monte Carlo
estimator in the central limit theorem for Markov chains \citep[see
e.g.][]{Asmussen2007}. In fact, lower ACTs will lead to smaller
time-averaged variance constants. Finally, we mention that in
practice, the target distribution is intractable and therefore the
variances of the components are unknown. However, these variances can
be estimated by the initial sequence estimator method.

Another measure of performance is the chain's average squared
jump distance (ASJD), which, for the $k$th component and repetition
$r$, we define as
\begin{equation*}
  \mathrm{ASJD}_{k}^{(r)} =\frac{1}{N}\sum_{i=1}^N|X_{i,k}^{(r)} - X_{i-1,k}^{(r)}|^2.
\end{equation*}
The average squared jump distance measures the movement of the chain
and also is linked with the acceptance rate of the MCMC method. Higher
values of average squared jump distances are desired as it indicates
larger moves and therefore more exploration of the space. We shall
consider the ASJD as a measure of a MCMC method to move around the
state-space. 

In summary, in the following results we are interested in the ACTs and
the ASJD per component. The distribution of the ACTs and ASJDs over
the repetitions, i.e.\ $\{\mathrm{ACT}_{k}^{(r)}:r=1,\dots,R\}$ and
$\{\mathrm{ASJD}_{k}^{(r)}:r=1,\dots,R\}$, will be presented using
violin plots.

\subsubsection{Mixture of Gaussians}
\begin{figure}[tbp]
  \centering

 \subfloat[Autocorrelation times]{  \includegraphics[width=0.95\linewidth]{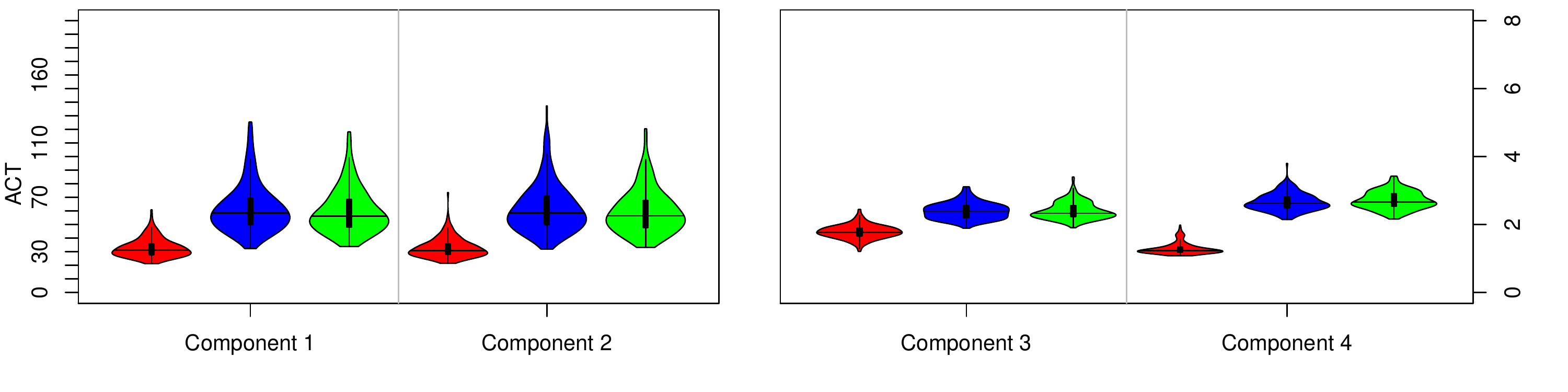}\label{fig:pi1act}}

 \subfloat[Average square jump distance]{
   \includegraphics[width=0.95\linewidth]{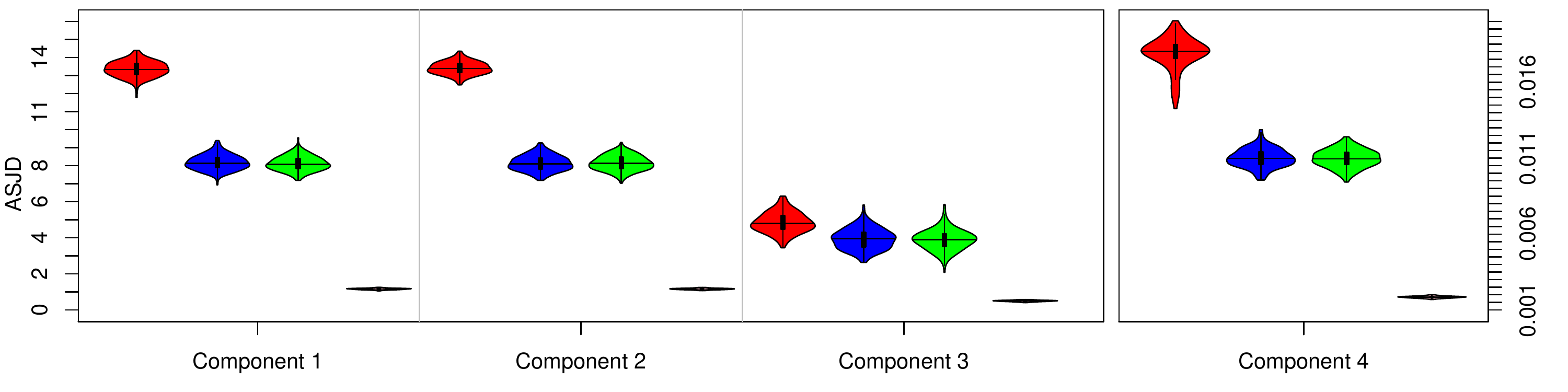}\label{fig:pi1asjd}
 }
  \caption{\textbf{Mixture of Gaussians, $\pi_1$:} Distribution of the ACTs and ASJDs of MCMC methods for target $\pi_1$. Red is AP, blue is AG1,
    green is AG2 and pink is MH.}\label{fig:pi1_res}
\end{figure}

The results for the 4-dimensional mixture of two Gaussians target,
$\pi_1$, are presented in Fig.~\ref{fig:pi1_res}.
Fig.~\ref{fig:pi1act} indicates that the AP method achieves lower ACTs
than the other methods for all components (including the MH method
which is not included in this figure due to very high ACTs). Further,
the range of ACT values suggest that the AP method consistently
produces MCMC chains with lower ACTs.

In terms of the movement of the MCMC chains, the AP method outperforms
the other methods for this target distribution. In fact, the
ASJDs presented in Fig.~\ref{fig:pi1asjd} show that the AP method
moving around the state-space in larger jumps than the other
methods. Since the AP and AG1 methods use the same weight function as
discussed in Section \ref{sec:gener-proc-comp} this advantage is due
to using the Plateau proposals in contrast to Gaussian
  proposals. These results suggest that the AP method is able to move
between the two Gaussians in the target density efficiently.

\subsubsection{Banana Distribution}
\begin{figure}[tbp]
  \centering

\subfloat[Autocorrelation times]{ \includegraphics[width=0.95\linewidth]{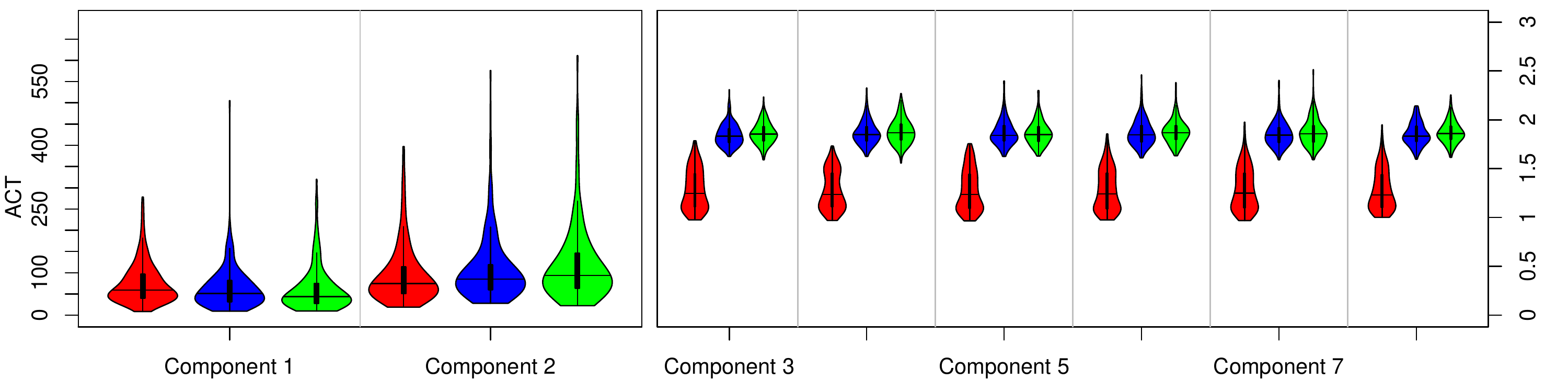}\label{fig:pi2act}}

 \subfloat[Average square jump distance]{
  \includegraphics[width=0.95\linewidth]{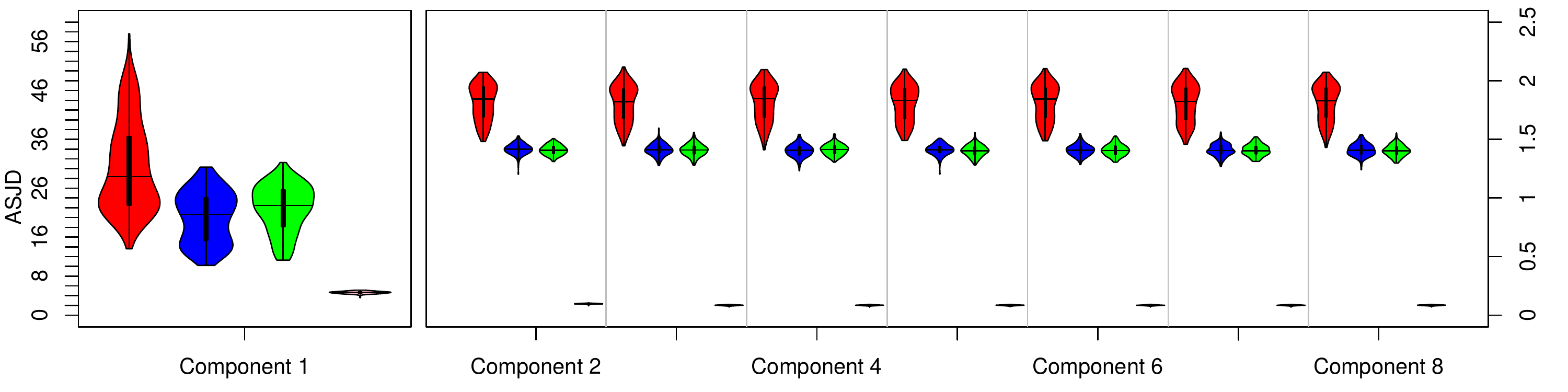}\label{fig:pi2asdj}
 }
  \caption{\textbf{Banana Distribution, $\pi_2$:} Distribution of the ACTs and
    ASJDs of MCMC methods for target $\pi_2$. Red is AP, blue is AG1,
    green is AG2 and pink is MH.} \label{fig:p2_res}
\end{figure}

The target, $\pi_2$, is a difficult distribution to sample from due to
the wide ranging variances in each component and the unusual
banana-shape of the first two components. 
The ACTs for the methods, presented in Fig.~\ref{fig:pi2act}, show
similar results across the AP, AG1 and AG2 for the first two
components. However, the remaining components the AP is achieving
notably smaller ACTs. The ACT results for the MH are substantially
larger for all components and thus are not included in this figure. As
an indication, the median ACTs for components 1 to 8 respectively are:
1131.74, 2066.35,   54.24,   54.37,   54.34,   54.03,   54.74,   54.47
for the MH
method to 2 decimal places.

The ASJDs for the methods presented in Fig.~\ref{fig:pi2asdj}. The AP
method again outperforms the other methods by achieving higher ASJDs
for all components. Note that for the first component the wide range
of jumping distance produced when using the Plateau proposals. This
suggests that the AP method is able to navigate the banana-shape in
the first component easily.

\subsubsection{2D Perturbed Distribution}
\begin{figure}[tbp]
  \centering

 \subfloat[Distribution of log autocorrelation times]{  \includegraphics[width=0.95\linewidth]{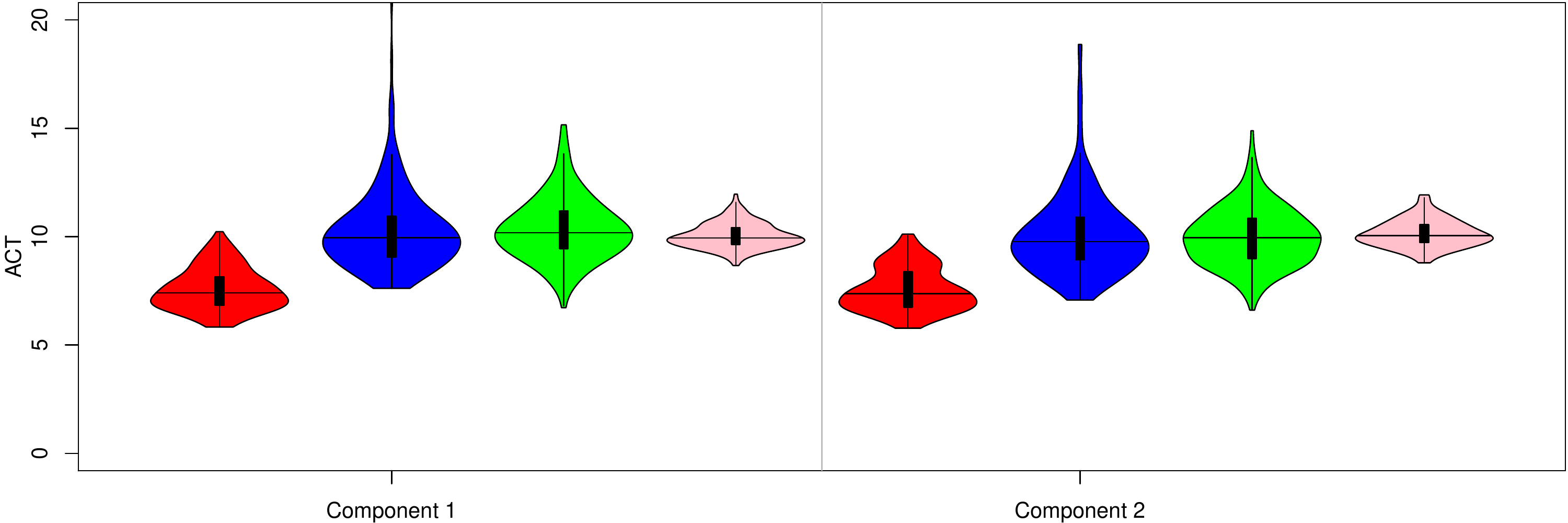}\label{fig:pi3act}}

\subfloat[Distance of average square jump distance]{
    \includegraphics[width=0.95\linewidth]{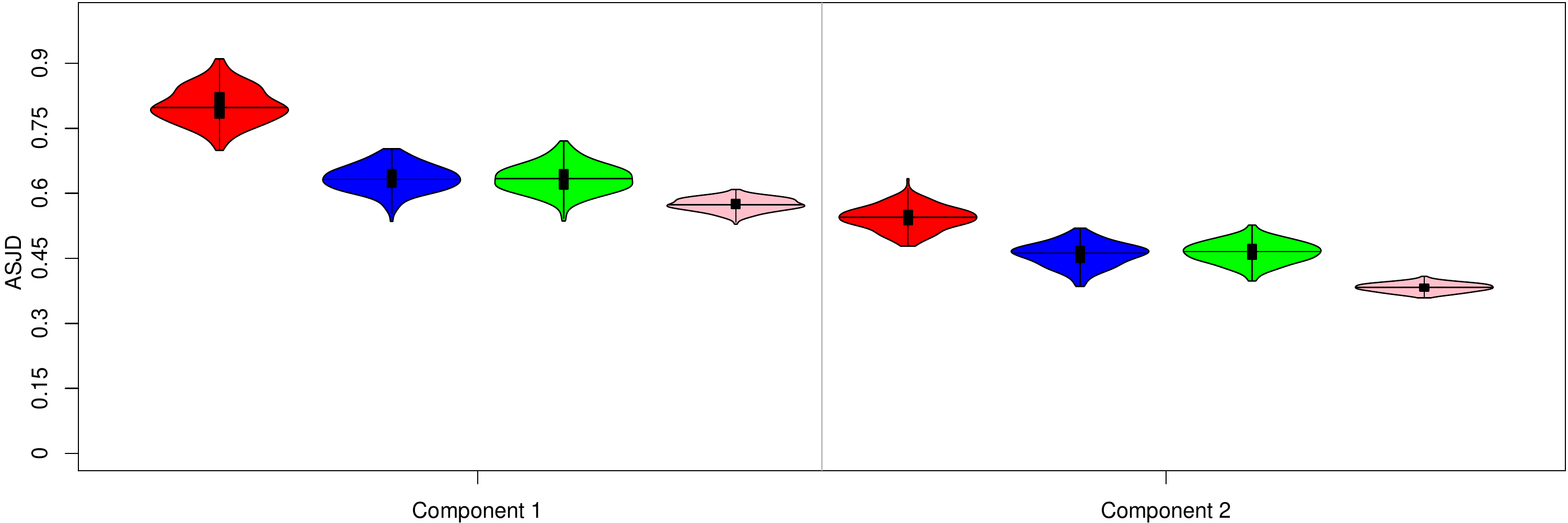}\label{fig:pi3asjd}
 }
  \caption{\textbf{2D Perturbed Distribution, $\pi_3$:} Distribution of the ACTs and
    ASJDs of MCMC methods for target $\pi_3$. Red is AP, blue is AG1,
    green is AG2 and pink is MH.}\label{fig:p3_res}
\end{figure}

For the perturbed 2-dimensional distribution, $\pi_3$, the
perturbations represent local modes where MCMC methods may potentially
get stuck. Again, the AP method's ability to move slightly larger
distances, as shown in Fig.~\ref{fig:pi3asjd}, gives it a slight
advantage over the other methods. This ability to jump further may
explain the lower ACTs for the AP method as depicted in Fig.~\ref{fig:pi3act}.

\subsubsection{1D Perturbed Distribution}
\begin{figure}[tbp]
  \centering
 \subfloat[Distribution of autocorrelation times]{  \includegraphics[width=0.45\linewidth]{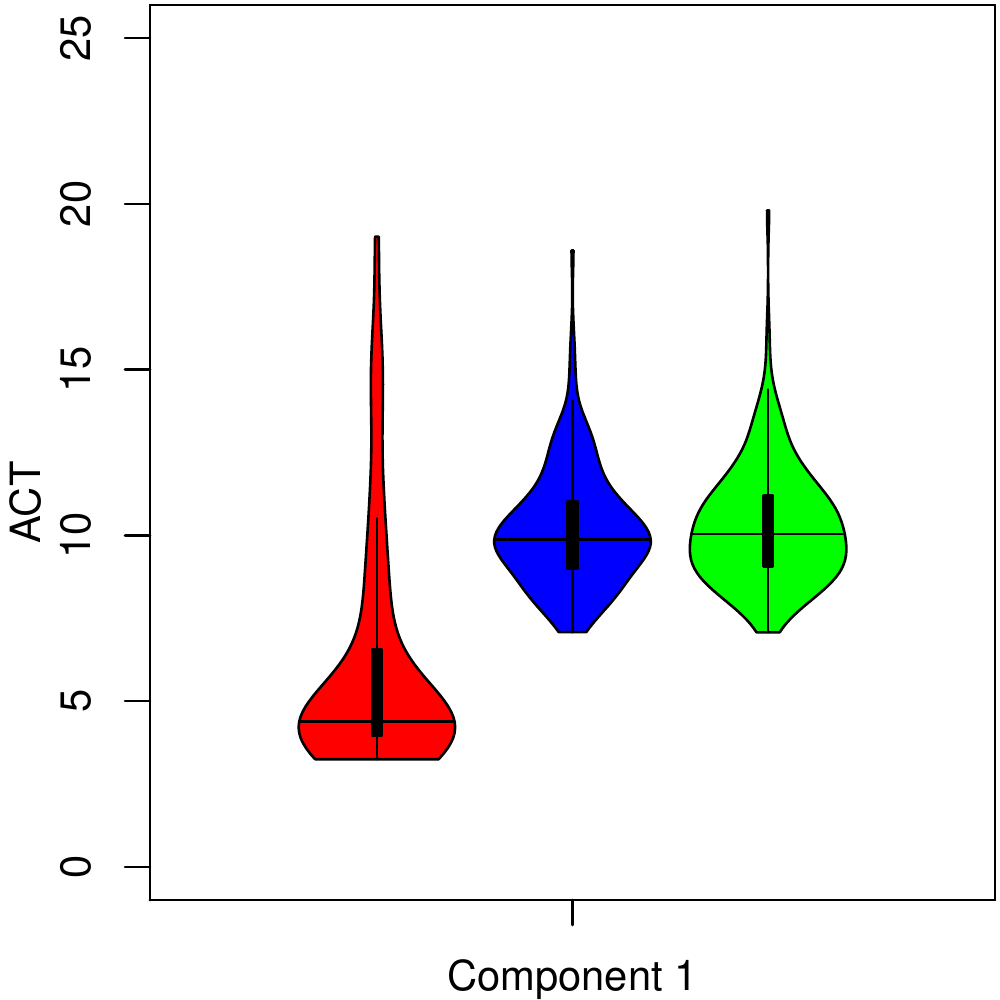}\label{fig:pi4act}}
\subfloat[Distribution of average square jump distance]{    \includegraphics[width=0.45\linewidth]{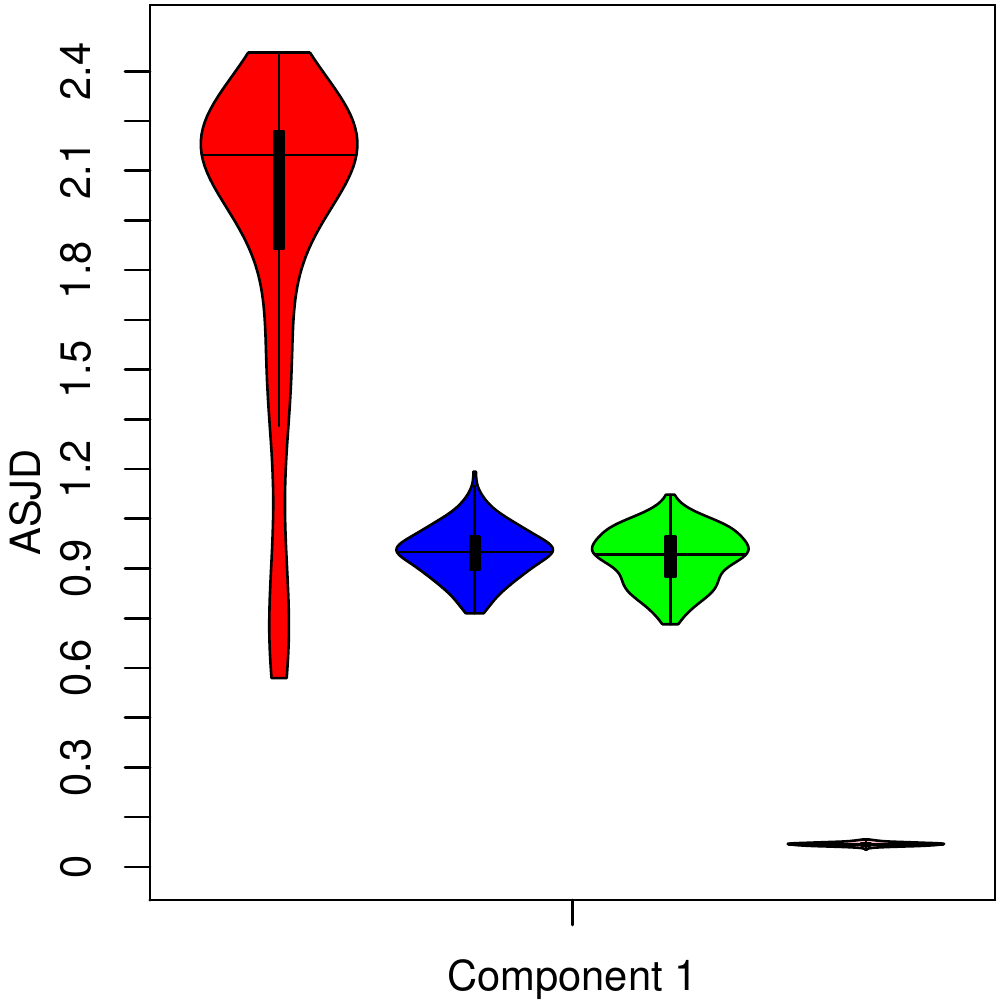}\label{fig:pi4asjd}}
  \caption{\textbf{1D Perturbed Distribution, $\pi_4$:} Distribution of the ACTs and
    ASJDs of MCMC methods for target $\pi_4$. Red is AP, blue is AG1,
    green is AG2 and pink is MH.}
  \label{fig:p4_res}
\end{figure}

Similar to $\pi_3$, the oscillations in $\pi_4$ are potential areas
where an MCMC may get stuck. The ACTs for the AP, AG1 and AG2 method
are presented in Fig.~\ref{fig:pi4act}. The AP achieves the lowest
ACTs, however there are a few outliers which may indicate a few runs
where the sampler got stuck in the local modes. This may also be the
case for the AG1 method. For the MH method, the ACTs (not presented in
the figure) are extremely large in comparison to the the other methods -- with a
median of 178.54 and a range of $(111.62, 457.56)$ to 2 decimal places. 

The ASJDs for the AP method, on average is jumping also twice the
distance of the AG1 and AG2 methods -- see
Fig.~\ref{fig:pi4asjd}. Again there are some outlying ASJDs for the AP
method which may indicate some repetitions where the MCMC got stuck in
local modes.

\section{Conclusion}\label{sec:conclusion}

In this paper we have introduced Plateau distributions as a novel-class of proposal distributions for the use with component-wise,
multiple-try Metropolis MCMC. These proposal distributions are a
combination of uniform distributions, leading to a family of
distributions with non-overlapping supports. The notion of using
non-overlapping proposals in multiple-try MCMC methods is intuitive
and, in fact, motivated as means to counter the disadvantages
(e.g.~inefficient proposing of trials) of greatly overlapping proposal
distributions such as Gaussians. Moreover, the class of Plateau
distributions are simple to implement for use as proposals in MCMC
methods and are straightforwardly combined with the simple, yet highly
effective, adaptation procedure presented in Section
\ref{sec:adaptation-proposals}. As mentioned in the introduction,
  the novelty of this work lies in both the Plateau proposals and the
  bespoke adaptation method. The designed adaptation method takes
  advantage of the non-overlapping proposals to better explore the
  space and ``scale'' the proposals to the target distribution. The advantages of
 our proposed algorithm over Gaussian proposals with a similar adaptation
  method was presented in simulations in Section
  \ref{sec:invest-adapt-mcmc}.

We have demonstrated that using the Plateau proposal distributions
with the suggested adaptation leads to MCMC methods that perform well
for a variety of target distributions. In fact, the results indicate
that using our method produces MCMC chains that explore the
state-space better with lower autocorrelation times, when compared to other
adaptive multiple-try methods with greatly overlapping Gaussian
proposals. Furthermore, the simulation results suggest that the
Plateau proposals are able to efficiently sample from target
distributions with distance modes, complex shapes, and many nearby
modes.

The results and the simplicity of their design makes the Plateau
proposals appealing for general use in component-wise, multiple-try
MCMC algorithms. As a matter of fact, the introduced class of Plateau
distributions is one type of non-overlapping proposals. Further
research may investigate other types of non-overlapping proposals
which may have multiple interacting trials (e.g.\ see
\cite{Casarin2013}) and may be asymmetric. Further theoretical
research is required to determine the mixing properties of the MCMC
chain produced by these Plateau proposals and adaptation procedure.

\newpage
  \appendix
  \section{A note of convergence of adaptive component-wise multiple-try
    algorithms}\label{sec:note-conv-adapt}
The convergence (in total variation distance) of algorithms
of the form of Algorithm \ref{alg:very_generic} described in Section 4 has been proven in
\cite{doi:10.1080/10618600.2018.1513365}.

The proof of convergence is ensured by the algorithm, both the MCMC
algorithm and the adaptation procedure, satisfying two
conditions: diminishing adaptation and containment. As mentioned
earlier, diminishing adaptation is satisfied by adapting with
probability $P_n = \max\left\{0.99^{n-1},1/\sqrt{n}\right\}$. For
containment to hold two technical, but not practical, modifications
are required -- these follow directly from
\cite{doi:10.1080/10618600.2018.1513365} and are presented as
quotations below with altered notation. The first modification is to
\begin{quotation}``
  ... choose a very large nonempty compact set $K\subset \mathcal{X}$ and
  force $\bs X_n\in K$ for all $n$. Specifically, we reject all proposals
  $\bs Y \notin K$ (but if $\bs Y\in K$, then we still
  accept/reject $\bs Y$ by the usual rule)...
  "
\end{quotation}
The second modification which is altered for our proposed Plateau
distributions is
\begin{quotation}``
  ... choose a very large constant $\Delta$ and a very small constant
  $\epsilon>0$ and force the proposal width $\delta$ to always be in
  $[\epsilon, \Delta]$...
  "
\end{quotation}
The proof then follows Section 3.5 in
\cite{doi:10.1080/10618600.2018.1513365}.

\end{document}